\documentclass[journal=jpcafh, manuscript=article, layout=twocolumn]{achemso}

\usepackage{amsmath, amssymb}%
\usepackage{subcaption}%
\usepackage{multirow}
\usepackage[version=4]{mhchem}%

\usepackage{xr-hyper}%
\usepackage{xr}%

\makeatletter
\newcommand*{\addFileDependency}[1]{
  \typeout{(#1)}
  \@addtofilelist{#1}
  \IfFileExists{#1}{}{\typeout{No file #1.}}
}
\makeatother

\graphicspath{}

\usepackage{color}

\title{
Low density interior in supercooled aqueous nanodroplets expels ions to the subsurface}%

\author{Shahrazad M.~A.~Malek}
\affiliation{
 Department of Physics and Physical Oceanography, Memorial University of Newfoundland, 
 Canada, A1B 3X7
}

\author{Victor Kwan}
\affiliation{
Department of Chemistry, The University of Western Ontario, London, Ontario, Canada N6A 5B7
}

\author{Ivan Saika-Voivod}
\affiliation{
 Department of Physics and Physical Oceanography, Memorial University of Newfoundland, 
 Canada, A1B 3X7
}
\alsoaffiliation{Department of Applied Mathematics, Western University, London, Ontario, Canada, N6A 3K7}

\author{Styliani Consta}
\email{sconstas@uwo.ca}
\affiliation{
Department of Chemistry, The University of Western Ontario, London, Ontario, Canada N6A 5B7
}

\begin{document}

\begin{abstract}

The interaction between water and ions within droplets plays a key role in 
the chemical reactivity of atmospheric and man-made aerosols.
Here we report direct computational evidence that in supercooled aqueous nanodroplets  
a lower density core of tetrahedrally coordinated water expels the cosmotropic ions to the
denser and more disordered subsurface. In contrast, at room temperature, depending on the
nature of the ion  the radial distribution in the droplet core 
is nearly uniform or elevated towards the center. We analyze the spatial distribution of a single ion in terms of a 
reference electrostatic model.
The energy of the system in the analytical model is expressed as the sum of the 
electrostatic and surface energy of a deformable droplet. 
The model predicts that
the ion is subject to a harmonic potential centered at the droplet's center of mass. 
We name this effect ``electrostatic confinement''.
The model's predictions are consistent with the simulation findings for a single ion 
at room temperature but not at supercooling. 
We anticipate this study to be the starting point for investigating the
structure of supercooled (electro)sprayed droplets that are used to preserve
the conformations of macromolecules originating from the bulk solution.

\end{abstract}

\maketitle

\section{\label{sec:intro}Introduction}

The spatial distribution of ions in droplets plays a decisive role in chemical reactivity in
atmospheric and man-made aerosols.
Applications of the man-made aerosols relevant to this study include spray-based ionization methods
used in native mass spectrometry analysis\cite{raab2020evidence} and use of droplets as 
micro- (nano-) reactors for accelerating chemical synthesis\cite{cooks2015, zare2016, sahraeian2019droplet}.
Aerosol droplets in the lower atmosphere carry a small charge
determined by at most a few excess ions whereas droplets in 
thunderclouds and (electro)sprays are highly charged.
In this article we show how a heterogeneous solvent structure in supercooled aqueous mesoscopic clusters 
charged with ions changes their spatial
distribution relative to that at room temperature. 
Hereafter, we will use the term nanodroplets for these mesoscopic clusters.

The structure and stability of supercooled clusters composed of ionic species 
has fascinated scientists over several decades. Most of the experiments have
been performed for clusters composed of a few tens of water molecules. Experiments have 
detected abundance of certain ``magic''  
cluster sizes associated with clathrate structures\cite{lee1998freeze} and have studied 
their reactivity in atmospheric chemistry\cite{schindler1996heterogeneously}.
Many of these experiments have used Fourier transform ion cyclotron resonance (FT-ICR) mass spectrometry where
dominant evaporative cooling of clusters over heating due to the absorption of black body radiation 
from the warmer walls of the apparatus \cite{lee1998freeze, schindler1996heterogeneously} has 
been reported. 
In certain FT-ICR experiments the temperature of clusters composed of 50-70\ce{H2O} molecules is 
estimated to be 130~K-150~K. 
The hydration of ions in clusters at supercooling and elevated
temperature in the last decade is actively investigated using infrared 
photodissociation spectroscopy\cite{pradzynski2012fully, stachl2020effects, cooper2016delayed}.
Moreover, supercooled droplets are used to preserve peptide conformations 
originating from the bulk solution in ion mobility and mass spectrometry experiments\cite{raab2020evidence}.

Most of the computational studies\cite{lu1996ion, gorlova2016characterization, herce2005surface, perera1993ion,  
  thaunay2017dynamics, zhao2013first, hagberg2005solvation, burnham2006properties, 
  makov1994solvation, fifen2016structure, galib2017revisiting, 
  perera1992structure, perera1993erratum,
perera1991many, hagberg2005solvation, caleman2011atomistic, 
werhahn2014universal, vaitheeswaran2006hydrophobic, 
harder2008origin,herce2005surface,galib2017revisiting, fracchia2018force, kwanJASMS2020, duignan2020method} 
are for clusters at room or elevated  
temperature, 
while there are relatively few computational studies\cite{burnham2006properties,zhao2013first}
of the supercooled clusters due to fact 
that they are notoriously challenging to be equilibrated. 
Voth and co-workers\cite{burnham2006properties} have investigated the location of a single \ce{Na+}, \ce{Cl-} and
\ce{H3O+} ion in clusters of up to 100 \ce{H2O} molecules
in the temperature range of 100~K-450~K.  
They found that in supercooled clusters of 100 \ce{H2O} molecules
\ce{H3O+} and \ce{Na+} tend to reside within a few monolayers of the
surface. At room and elevated temperature both the \ce{Na+}
and \ce{Cl-} ions are found nearer to the center, while the \ce{H3O+} continues to
be near the surface. 

The spatial distribution of multiple ions in nanodroplets has been 
investigated less\cite{zdrali2019specific,kwan2019ions,kwan2020bridging}. 
Previously\cite{kwan2019ions,kwan2020bridging}  
we reported atomistic simulations of the location of  multiple ions 
in aqueous nanodroplets with diameter  $\approx 2$~nm - 16~nm at a temperature 
range of 300~K to 450~K. It was found that in droplets comprising 
$\approx 1000$ \ce{H2O} molecules, the radial ion distribution 
(measured from the droplet's center of mass - COM) is almost uniform. 
As the droplet size increases, the distribution shows a distinct maximum in the outer droplet layers. The distribution decays almost exponentially toward the droplet's COM.
Toward the droplet exterior, the decay is determined by  the ion size
and shape fluctuations\cite{kwan2019ions}. The solution of the non-linear Poisson-Boltzmann 
equation for a spherical geometry\cite{malevanets13variation} is used as a reference model to compare with the atomistic simulations. The distributions of multiple ions and biological molecules such as peptides in supercooled droplets are still completely unknown.

The related question of the propensity of ions in a liquid-vapor planar interface has
been studied over several decades\cite{enami2013long, enami2014ion, jungwirth2000surface, 
omta2003negligible, ghosal2005electron, knipping2000experiments, 
otten2012elucidating, petersen2006nature, onorato2010measurement, 
smith2017soft, baer2011toward, adel2021insight, beck2013influence, tobias2008getting, tielrooij2010cooperativity}.
It is emphasized that the forces that determine the location of multiple ions in a planar
interface are not the same as those in charged droplets regardless of their size.
The reason is that the conductivity of the aqueous droplets drives the excess ions 
nearer to the surface and the counterions toward the interior\cite{kwan2020bridging}. 
Another difference arises from the type of fluctuations that a droplet undergoes.
Typical fluctuations in a charged
droplet, regardless of its size, are conical protrusions on the surface, which
are absent in a planar interface\cite{consta2003fragmentation, kwan2021relation}.
These fluctuations may lead to release of ions from droplets.
A single ion in a droplet is always subject to a harmonic potential centered to the
droplets' COM, as we show in this article. Such a force is not present
in a planar interface. Detailed chemical
interactions (e.g. charge transfer, solvent network) add one more layer
of complexity in determining the ion position.

Here we study the radial distribution profiles of a single ion and multiple ions in relation to the
solvent organization in supercooled aqueous clusters with sizes that vary from 100 \ce{H2O} molecules (corresponding to a diameter of $\approx$ 1.8~nm) to 1100 (diameter $\approx$ 4.0~nm). 
Our recent computer simulations~\cite{maleknc2018} for pristine supercooled water clusters
have revealed an anomalous, inverted radial density profile emerging for $N\ge200$ molecules at low temperature, in which a low-density core with relatively good 
tetrahedral ordering~\cite{malek2019} is surrounded by a higher density subsurface.  
These observations have some commonalities with studies of nucleation within droplets and 
thin films\cite{johnston2012crystallization, nandi2017ice, li2013ice, 
haji2017computational, hall2018does}. 
We hypothesize that the low density water core
in supercooled droplets will differentiate the ion distribution from that
at room temperature. 
This hypothesis has not been explored earlier\cite{burnham2006properties}. 

In order to obtain insight into the fundamental forces that determine the
location of a single ion, we 
introduce a reference analytical model of 
a fluctuating droplet that contains a single ion.  
To our knowledge, such a reference model 
is still missing from the literature. In the model
the energy of a fluctuating droplet is the sum of the electrostatic energy  and surface energy.
It predicts that an ion is always subject to a harmonic potential 
centered at the droplet's COM. For this reason we name this effect ``electrostatic confinement'' (EC).
The EC effect competes with the geometric confinement.
We present the conditions under which the electrostatic confinement is more evident.

\section{Theory of Electrostatic Confinement}
Here, we present the key points of an analytical model that provides insight into the forces
that determine the location of a single (macro)ion in a droplet. The details of the model
are found in Sec.~S1 in the SI.
In the model we consider a charged dielectric droplet that may undergo shape fluctuations.
The charge carrier can be a simple ion such as a \ce{Na+} or a single  macroion
with bound charge such as a protein or a nucleic acid.
The energy of the charged droplet is written as the sum of the electrostatic
energy and surface energy. 

\begin{figure}[htb!]
  \includegraphics[width=\linewidth]{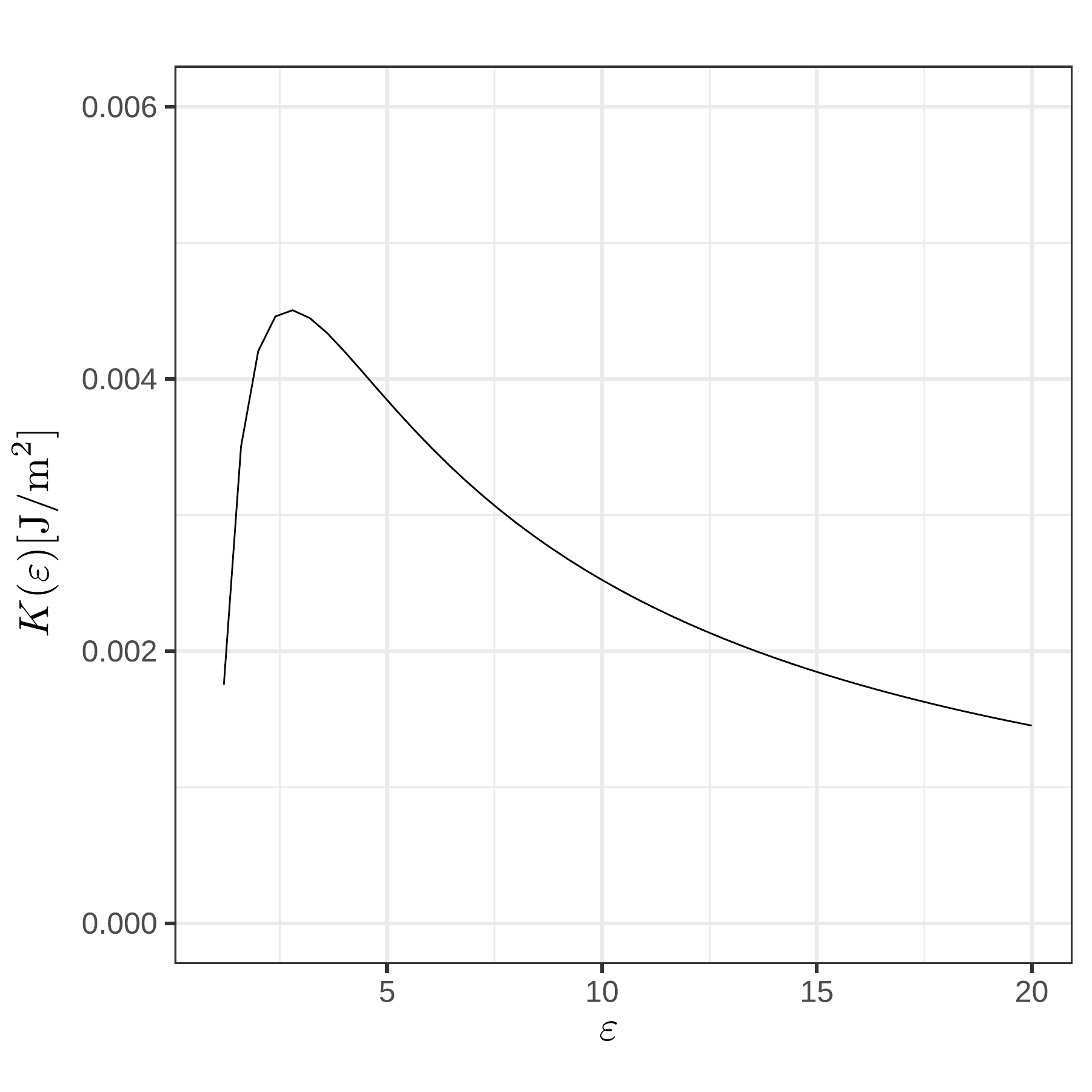}
  \caption{Magnitude of the spring constant (Eq.~\ref{eq:Ke}) as a function of the
    relative dielectric constant $\varepsilon$. The values
    correspond to an ion of charge $Q=1e$ in a droplet comprising 1000
    water molecules and radius 19\AA. The value of the potential
    has a maximum at ${\varepsilon\approx 1+\sqrt{3}\approx 2.73}$. }
  \label{fig:kappa}
\end{figure}

After some algebra (details are found in Sec.~S1 in SI), 
it is found that the energy (denoted by $\Delta E_1$) related to the distance of the ion
from the droplet's  COM is given by
\begin{equation}
\Delta E_1(\|\mathbf{r}\|) =
  \frac{\varepsilon-1}{4\pi\epsilon_0\varepsilon(\varepsilon+2)}	
  \frac{Q^2}{R^3}
  \|\mathbf{r}\|^2 
  \label{eq:ener-ion}
\end{equation}
where $Q$, $R$ and $\varepsilon$ are the charge of the ion, the
droplet radius and the relative dielectric constant of the solvent,
respectively, $\epsilon_0$ is the vacuum permittivity
and $\|\mathbf{r}\|^2 = X^2_{\mathrm{COM}} + Y^2_{\mathrm{COM}}  +Z^2_{\mathrm{COM}}$ 
(where $X_{\mathrm{COM}}, Y_{\mathrm{COM}}, Z_{\mathrm{COM}}$ are the coordinates of the droplet's COM).
The energy (\ref{eq:ener-ion}) has the functional form
of a harmonic potential. From this point we would refer to this effect
as ``electrostatic confinement'' (EC). We
introduce the spring constant ${K(\varepsilon)}$ where
\begin{equation}
  K(\varepsilon) = 
  \frac{\varepsilon-1}{4\pi\epsilon_0\varepsilon(\varepsilon+2)}	
  \frac{Q^2}{R^3} .
  \label{eq:Ke} 
\end{equation}
In Fig.~\ref{fig:kappa} we plot the value of the spring constant as a
function of the relative dielectric constant for a droplet comprising
1000 water molecules and an ion. The electrostatic energy has two
limiting cases ${\varepsilon=1}$ and ${\varepsilon=\infty}$ when the
electrostatic interaction of the ion with the droplet surface
vanishes. In the former case the external and internal dielectric
constants are equal and the droplet does not perturb the electric
field of the ion. In the latter case the electrostatic field is
localized in the vicinity of the ion and it is not affected by the
droplet surface. As seen in the plot the maximum of the coefficient
${K(\varepsilon)}$ is attained at
${\varepsilon=1+\sqrt{3}\approx 2.73}$.

\begin{figure}[htb!]
  \includegraphics[width=\linewidth]{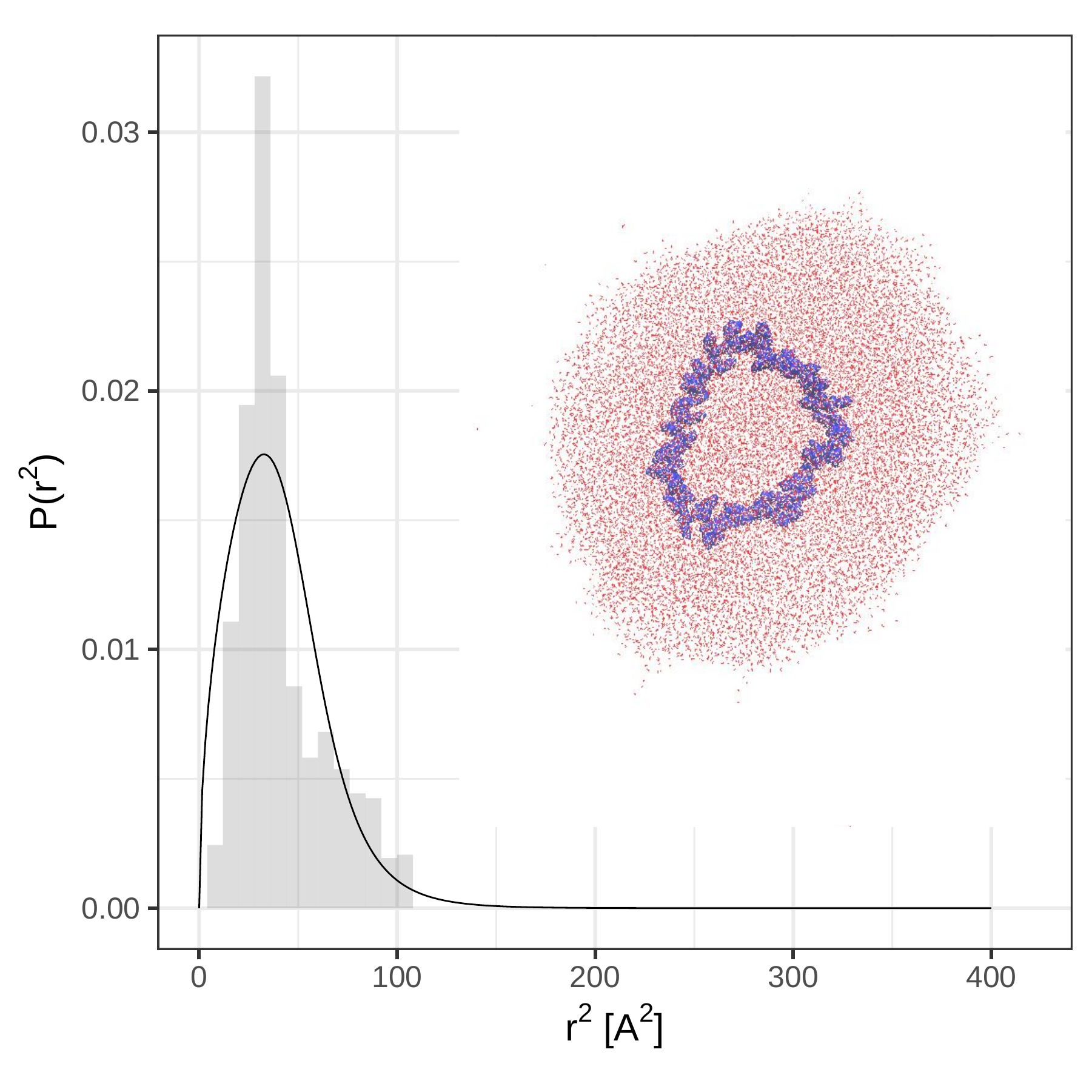}
  \caption{Distribution of the distances of the COM of
    the cyclic peptide (grey colored histogram) measured from the droplet's COM. The
    droplet radius is 3.8~nm and the charge of the peptide is 8\ce{e+}. The solid
    line is the gamma function fitted to the distribution using the
    maximum likelihood estimate (MLE). The inset shows
    a typical snapshot of the droplet composed of 8000 \ce{H2O} molecules (red colored) and
  the cyclic peptide (blue colored).}
  \label{fig:cyclic}
\end{figure}

If the ion is localized in the droplet interior the Gibbs-Boltzmann
distribution of the ion positions is given by
\begin{equation}
  P(\|\mathbf{r}\|^2) = \frac{2}{\sqrt{\pi}}
  \left(\frac{K(\varepsilon)}{k_{B}T}\right)^{3/2} \|\mathbf{r}\|
  e^{-K(\varepsilon)\|\mathbf{r}\|^2/k_{B}T} . 
  \label{eq:gibbs}
\end{equation}
This probability distribution yields the radial concentration profile of the ion
along the droplet radius.
The expectation value of the square of the distance of the ion from
the center of mass is given by
\begin{equation}
  \langle \|\mathbf{r}\|^2 \rangle = \frac{3}{2} \frac{k_{B}T}{K(\varepsilon)}.
  \label{eq:r2-exp}
\end{equation}
The EC is more pronounced when the ion is localized at the center of
the droplet, therefore we can write
${\langle \|\mathbf{r}\|^2 \rangle/R^2 \ll 1}$.  Analyzing
Eqs. (\ref{eq:Ke}) and (\ref{eq:r2-exp}) we conclude that the effect
will be more pronounced at low temperature, high charge, small radius
and intermediate values of the dielectric constant. Small droplet with
high charge may undergo Rayleigh
instability\cite{rayleigh1882,peters1980rayleigh, hendricks1963, consta2015disintegration} 
leading to the ``star''-shaped droplets\cite{consta2010manifestation} (see Fig.~S2 in SI). 
If this is an issue in
observations we need to increase the droplet radius while keeping
constant the value of the Rayleigh parameter ${X \sim Q^2/R^3}$\cite{kwan2021relation}. To
illustrate the EC we model high charges by creating models of charged
cyclic peptides. We could have used in the simulations a spherical ion with charge 8\ce{e+} 
(or any other charge) instead of a cyclic peptide. The cyclic peptide is preferred 
because it is an effective way to
create a large compact charge, which
gives rise to realistic local electric fields
that are not high enough to cause water dissociation.
In experiments, a single ion with a high charge (e.g. 8\ce{e+}) may give rise to
high local electric fields that may lead to water dissociation. The distribution of
charge in a cyclic peptide does not cause this problem.
DNA and RNA strands are other examples where the
effect of the EC will be clearly observed. When a linear nucleic acid is used 
the droplet size should be large enough so that
the linear geometry of the macroion does not affect
the droplet's spherical shape.

Using the maximum value of the $K$ parameter from Fig.~\ref{fig:kappa}
we obtain the estimate of the minimal dimensions of the excursions of
the ion from its equilibrium position at the center of the droplet
${\sqrt{\langle \|\mathbf{r}\|^2 \rangle} \geqslant \mathrm{12\AA} }$.
Therefore, for a droplet comprising 1000 water molecules and a single
charge ${Q=1e}$ the geometric confining effects should be taken into
consideration.

In Fig.~\ref{fig:cyclic} the distribution of the distance of the center
of mass of the peptide relative to the COM of the droplet is
plotted. The droplet's equimolar radius $(R_e)$ is 3.8~nm and the charge is 8\ce{e+}. The
distribution tapers off before reaching the droplet surface. Sampling
proves to be a challenge in such systems.  The simulation time should
be much longer than the time for a molecule to diffuse the width of
the droplet ${D t_{\mathrm{sim}} \gg R^2 }$. Typical values of the
diffusion coefficient ($D$) are ${\sim 10^{-9} \mathrm{[\frac{m^2}{s}]} }$,
hence the simulation time has to be in 10~ns~100~ns range at temperature
${T=\mathrm{300K}}$.

In summary, we showed that there is always a force on an ion toward the droplet's COM. The EC 
effect is more evident for an ion with a charge of at least $\pm 3e$ (where $e$ is
the elementary positive charge) found
in a droplet with a relatively small radius (Fig.~S1 in SI). For the 
EC to become evident the radius size will be equal or moderately larger than
the radius at the Rayleigh limit\cite{rayleigh1882, hendricks1963, consta2015disintegration}.
The Rayleigh limit is defined as the point where the electrostatic forces balance
the surface forces.
For a radius smaller than that at the Rayleigh limit, the
droplet shows instability that takes ``star''-shapes (Fig.~S2 in SI).
The star-shapes are minimum energy structures.
The EC model can be extended further by including other factors such as the ion size and 
hydrophobic effects.

\section{Experimental: Models and Simulation Methods}
Here we present the main points of the computational methods. A detailed account of the
methodology and parameters are found in Sec.~S2 in SI. 
We performed equilibrium molecular dynamics (MD) simulations of aqueous
nanodroplets charged with a single ion \ce{Na+}, \ce{Li+}, \ce{Cs+}, \ce{F-}, \ce{Cl-}, \ce{I-}, and multiple \ce{Na+} ions.
The droplets range in size from 100 \ce{H2O} molecules 
($R_e = 0.87$~nm) - 1100 H2O molecules ($R_e = 2.0$~nm).
The temperature of the droplets was set at $T=200$~K for supercooled droplets and
at 300~K (room temperature) except for the 100-\ce{H2O}-molecule droplets that were simulated at
260~K instead of 300~K.
The MD simulations were performed using
 GROMACS v4.6.1~\cite{Berendsen1995,Lindahl2001,van2005,Hess2008} and NAMD v2.14\cite{phillips05scalable}.
The water molecules are modeled with the TIP4P/2005 (transferable intermolecular
potential with four points) model~\cite{Abascal2005}. We selected the TIP4P/2005
because it reproduces well
the density anomaly in liquid water~\cite{Abascal2005,gonzalez2016}, the liquid-gas surface tension over a broad range of temperatures~\cite{vega2007} and the liquid-gas coexistence line in the density-temperature plane~\cite{vega2006}.
We also performed a number of simulations with a Drude oscillator-based
polarizable model, where the \ce{H2O} molecules are represented with the
SWM4-NDP model\cite{Lamoureux2006} and the ions, \ce{Na+} and \ce{Li+}
with the CHARMM Drude force field\cite{Yu2010,Luo2013}.
The equations of motion for the TIP4P/2005 set of simulations were integrated with a time step of 
2.0~fs and for
the SWM4-NDP model with 1.0~fs.
Each nanodroplet was placed in a closed volume so that it is in equilibrium with its
vapor. The volume is large enough to accommodate the droplet's shape fluctuations.
The simulations were carried out in the canonical ensemble -- constant number of 
molecules ($N$), volume, and $T$. 
The trajectories were visualized using VMD 1.9.4a47\cite{humphrey96vmd}.

To analyze the features of the radial distribution (concentration) profiles of the ions
in relation to the water structure, 
we use several parameters: (a) the density of water as determined from the Voronoi volumes associated with each water molecule,
denoted by $\rho_{V}(r)$ (where $r$ denotes the distance from the droplet's COM); (b) the tetrahedral order parameter, denoted by $q_T(r)$, and (c) the distance to the fifth nearest neighbor O of a given O atom denoted by $d_5$.

The parameter $q_T(r)$ is defined as follows.
Initially, the tetrahedral order parameter~\cite{qt}, at the level of a single particle 
is defined as,    
\begin{eqnarray}
q_i &=& 1-\frac{3}{8}\sum_{j=1}^3\sum_{k=j+1}^4 \left[\cos\psi_{jik}+\frac{1}{3}\right]^2, \label{qiT}
\end{eqnarray}
where $\psi_{jik}$ is the angle between 
an oxygen
atom $i$ and its nearest neighbor oxygen atoms $j$ and $k$ within a distance of $r_{cut}=0.35$~nm. 
The radial function $q_T(r)$ as
the average value of $q_i$ for all molecules within a spherical shell enclosed within $r\pm \Delta r/2$, where $\Delta r=0.05$~nm. Similarly, we report results for $d_5(r)$, the average distance to the fifth O neighbor for O atoms located in the same spherical shell centered at $r$.  
We note that we find sometimes slight, but no significant differences for $\rho(r)$, $\rho_v(r)$, $q_T(r)$ and $d_5(r)$ whether an ion is present or not, and therefore use well-sampled data for these quantities for pure water taken from previous studies~\cite{maleknc2018,malek2019}.

\section{Results}

\subsection{Single ion}

\begin{figure}[htb!]
    \centering
   \includegraphics[width=\columnwidth]{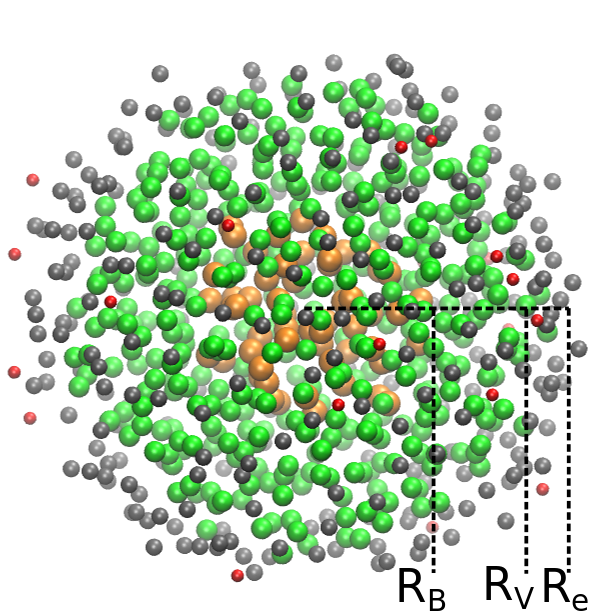}
    \caption{
    Regions in a typical snapshot showing O atomic sites of a system composed 
    of 776~\ce{H2O} molecules-\ce{Na+} at 200~K.
    From the center of mass to a distance $r = R_B$, the core structure (orange-colored) 
    is characterized by a bulk-like, low-density tetrahedral network.  The subsurface 
    region $R_B < r < R_V$ (green-colored), is characterized by a higher density, 
    lower tetrahedrality, and structure that changes with $r$.
    Surface atoms (grey-colored) in $R_V < r < R_e$ have a large Voronoi volume. The remaining
    atoms (red-colored) are found in $r> R_e$, where $R_e$ is the equimolar radius. 
    The determination of $R_B$ and $R_V$ is illustrated in Fig.~\ref{fig:N776} and discussed in the text.  
    The lower-density core and higher-density subsurface is a structural feature 
    emerging at low $T$ and  $N \ge 200$.
   }
    \label{fig:schematic}
\end{figure}

The main idea of the present article is that the low-density tetrahedral network
that forms at sufficiently low $T$ in the core of sufficiently large nanodroplets
tends to expel certain ions to the relatively higher-density and more disordered subsurface.  
Fig.~\ref{fig:schematic} illustrates this structure for the case of an $N=776$ nanodroplet.
The density structure of the nanodroplet is approximately the same for the nanodroplet whether 
it contains a single, several or no ions.
Using a $N=776$-\ce{Na+} nanodroplet as an example,
it is found that at 200~K, the $0< r< R_\mathrm{B} = 7.4$~{\AA} region includes on 
average 57.14 oxygen sites yielding a number density $33.66\pm 0.15$~$\rm{nm}^{-3}$.
The subsurface, $R_B < r < R_V = 14.8$~{\AA} includes on average 410.62 O sites yielding a 
number density $34.56\pm 0.15$~$\mathrm{nm}^{-3}$. 
At 300~K, the density in the core is $34.38\pm 0.15$~$\rm{nm}^{-3}$ and in the subsurface
$34.14\pm 0.15$~$\mathrm{nm}^{-3}$, and thus the difference at 200~K is inverted and significantly larger 
than  at 300~K.

\begin{figure}[htb!]
    \centering
    \includegraphics[width=\columnwidth]{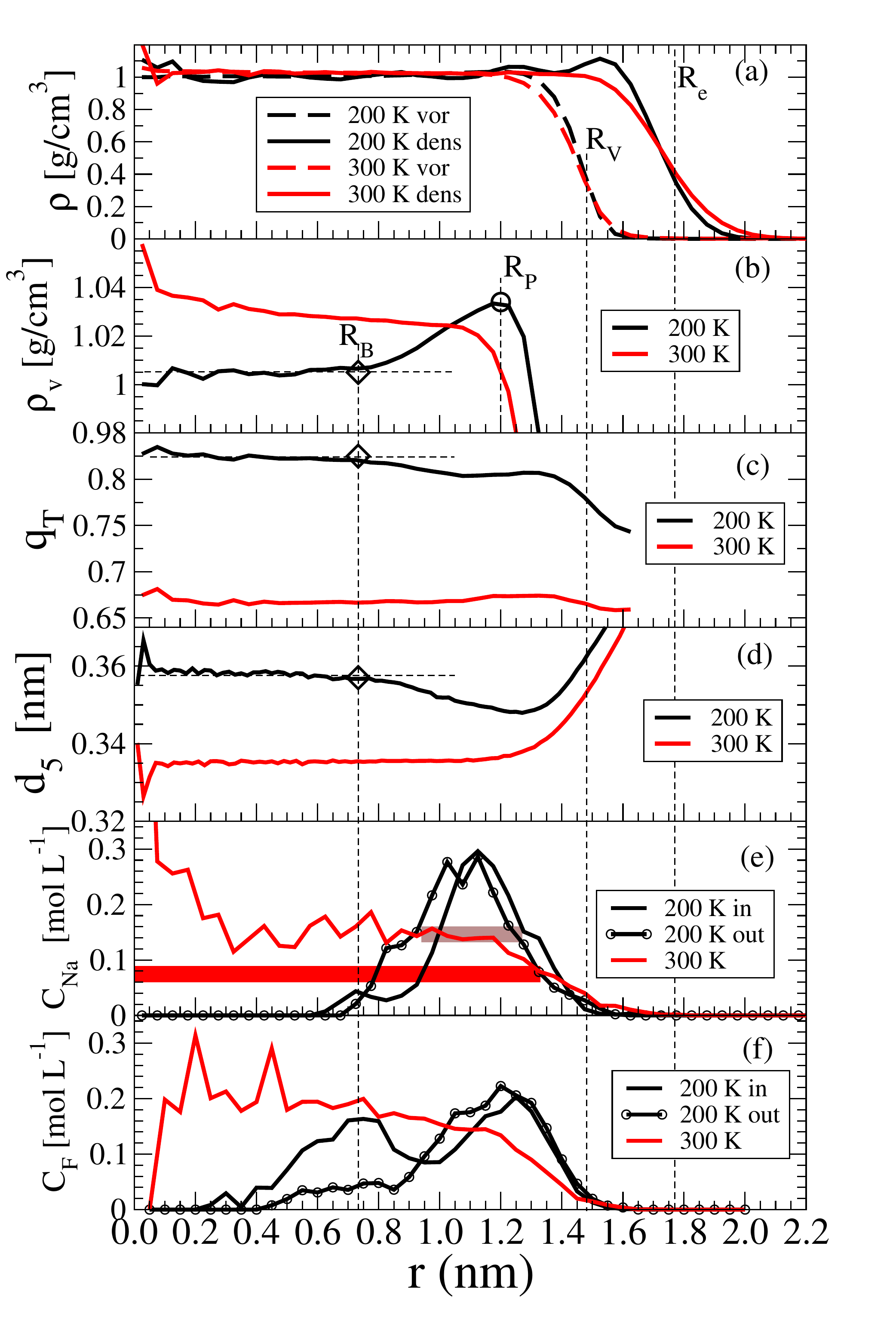}
    \caption{Structure of pristine water droplet, single \ce{Na+} and single \ce{F-}
    distributions for $N=776$ at $T=200$~K (black line) and $T=300$~K (red line).  Shown as functions of $r$ are 
    (a) density $\rho$ and density determined from Voronoi volumes $\rho_V$;  
    (b) a close up of $\rho_V$;
    (c) tetrahedral order parameter $q_T$; 
    (d) distance of the fifth O neighbor $d_5$;
    (e) \ce{Na+} concentration, $C_{\rm Na}$, for a single ion, which for 200~K includes results from starting the ion at the centre (in) and at the surface (out); 
    and
    (f) same as (e) but for \ce{F-}.
    In panel (b), $R_P$ marks the maximum in $\rho_V$, a feature representative of the subsurface at low $T$.
    The horizontal bars in panel (e) approximate the spatial extent of $C_{\rm Na}$ for 200~K (brown) and 300~K (red), and appear in Fig.~\ref{fig:Summary}. See details in the text.
    }
    \label{fig:N776}
\end{figure}

A more quantitative description is provided in Fig.~\ref{fig:N776}, which shows the relation
of the water structure and the probability density of the ion at 300~K (red curves)
and 200~K (black curves).  Fig.~\ref{fig:N776}~(a) shows the \ce{H2O} radial density ($\rho(r)$) 
using the physical volume.
Even though $\rho(r)$ profiles are typically used to identify a cluster's boundary, they may mask
certain features.
The decrease in the density profile close to the cluster's boundary
masks the role of the shape fluctuations. Also, the fact that the profiles are built
around the cluster's COM, which is not one of the molecular entities in the system, 
may overemphasize the layered structure of the solvent at supercooling. In larger particles (e.g. aerosols)
the plot will provide the average density
of the droplet as a function of radius and layering will not be present.
For this reason, we use additional measures of
the solvent structure such as the Voronoi volume (Fig.~\ref{fig:N776}~(a) and (b)), $q_T$ (Fig.~\ref{fig:N776}~(c)) and 
$d_5$ (Fig.~\ref{fig:N776}~(d)).
Note that for $r<0.2$~nm, good statistics are difficult to obtain for all radial quantities,
and results in this regime are quite noisy.
The density using the Voronoi volume ($\rho_V(r)$) is also shown in Fig.~\ref{fig:N776}~(a).
The main feature of $\rho_V(r)$ is that it decays rapidly for surface atoms -- those with
large Voronoi volumes that extend outside the nanodroplet.  To define a 
surface region we find $R_V$ such that $\rho_V(R_V)=\rho(R_e)$.  Given this, 
approximately two thirds of the molecules at $r = R_V$ have unbounded Voronoi volumes.
In Fig.~\ref{fig:N776}~(b) we show a close-up of $\rho_V(r)$.  While monotonically decreasing with $r$ 
at large $T$, at low $T$, $\rho_V(r)$ is approximately constant for $r< R_B$ and then increases to a maximum at $r=R_P$. We choose $r = R_P$ as a representative feature of the subsurface, 
which now by default occupies $R_B < r < R_V$.  Fig.~\ref{fig:N776}~(c) and Fig.~\ref{fig:N776}~(d), 
showing $q_T(r)$ and $d_5(r)$, 
respectively, likewise show a bulk-like core with uniform properties for $r < R_B$ and a 
more disordered subsurface.
The parameter $d_5$, a more indirect measure of the quality of the tetrahedral network,
shows a more disordered structure in the vicinity of the subsurface
density peak at low $T$.  
We note that $q_T(r)$ begins to increase beyond a minimum located at 1.6~nm, 
a location clearly in the surface layer where $d_5$ is quite high and $\rho_V$ is nearly zero.  
We thus cut off the $q_T(r)$ at this minimum, as the increase is not indicative of 
increased tetrahedral order.

Fig.~\ref{fig:N776}~(e) shows \ce{Na+} concentration $C_{\rm Na}(r)$ at 200~K and 300~K when 
a single \ce{Na+} ion is present in the nanodroplet.
At low $T$, $C_{\rm Na}(r)$ as obtained from starting the ion near the centre
and starting near the surface are compared -- thus showing the degree of equilibration we achieve. 
At $T=300$~K, $C_{\rm Na}(r)$ is approximately
constant in the interior of the droplet, and begins to decay within the subsurface, 
becoming small by $R_V$ and decaying to zero significantly before $\rho(r)$ does.  
At room temperature the \ce{Na+} concentration profile is consistent with the EC model. 
Because of the low charge, the geometric confinement effect dominates over the force
toward the droplet's COM. For this reason, the \ce{Na+} radial distribution 
does not distinctly peak at the droplet's COM, but it appears to be almost uniform 
with a slight increase toward the droplet's COM. 

Fig.~\ref{fig:N776}~(e) shows a significant difference in $C_{\rm Na}(r)$ at 200~K.
Rather than being almost uniform, $C_{\rm Na}(r)$ has a peak located within
0.2~nm from $R_P$.  Thus, we see that in a nanodroplet
with a heterogeneous radial density, as determined by $\rho_V(r)$,
the single \ce{Na+} ion tends to reside in the highest density environment for most of the time.
This tendency is consistent with the fact that for constant $T$ and polarization factor
(degree of dipole ordering)~\cite{Aragones2010}, the dielectric constant
increases with increasing density.

\begin{figure}[htb!]
    \centering
    \includegraphics[width=\columnwidth]{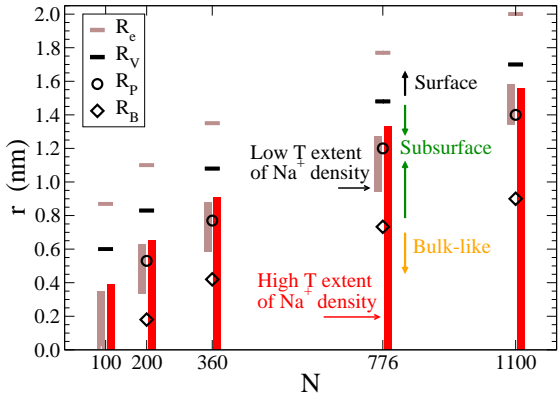}
    \caption{Radial features of nanodroplets comprising $N$ water molecules and  a single \ce{Na+}.  The features shown for each $N$ are determined as illustrated in Fig.~\ref{fig:N776} for $N=776$.}
    \label{fig:Summary}
\end{figure}

The thick horizontal bars shown in Fig.~\ref{fig:N776}(e) measure the approximate extents 
of $C_{\rm Na}(r)$ for low (brown) and high (red) $T$, and are drawn between 
$C_{\rm Na}$ values of half the peak or plateau height.  
For $300$~K, notwithstanding noise below $r=0.2$~nm, the distribution extends to 
the center of the droplet, and so the left end point of the red bar is at $r=0$.  
These horizontal bars are replotted vertically in Fig.~\ref{fig:Summary} 
along with $R_B$, $R_P$, $R_V$ and $R_e$, for other nanodroplet sizes. 
The detailed \ce{Na+} radial distribution profiles and the histograms of the raw data for the various droplet sizes
are shown in Figs.~S3 and S4, respectively, in SI.
Fig.~\ref{fig:Summary} shows the same trend in the structure of the \ce{H2O} and the 
location of the ions as found for $N=776$. Measures of structure for low
temperature ($T=200$~K) as those shown in Fig.~\ref{fig:N776} and 
in detail in Fig.~S4 in SI are summarized in Fig.~\ref{fig:Summary}.
The convergence of the trajectories for \ce{Na+} starting on the surface and the droplet's
COM is shown in Fig.~S5 in SI.
The surface thickness is estimated to be $R_e-R_V \approx 0.3$~nm and the subsurface thickness $R_V - R_B \approx 0.7 - 0.8$~nm for all sizes studied.  It is noted that the cluster of $N=100$ does not show a 
tetrahedrally organized core.

Similarly to \ce{Na+}, at 300~K, \ce{F-} (Fig.~\ref{fig:N776}~(f)) and \ce{Li+} (Fig.~S6 in SI)
show higher concentration in $0 < r < R_B$. 
Behavior consistent with our present study
for \ce{F-} at 300~K has been
reported in Ref.~\cite{petersen2006nature, zhao2013first} for clusters composed of 124-145~\ce{H2O} molecules.
However, these systems are too small to show the effect of a core and subsurface in the ion location.
We find that at 200~K, the maximum of the \ce{F-} concentration is shifted into the subsurface ($R_B < r < R_V$),
and similarly for \ce{Li+}.

\begin{figure}[htb!]
    \centering
    \includegraphics[width=\columnwidth]{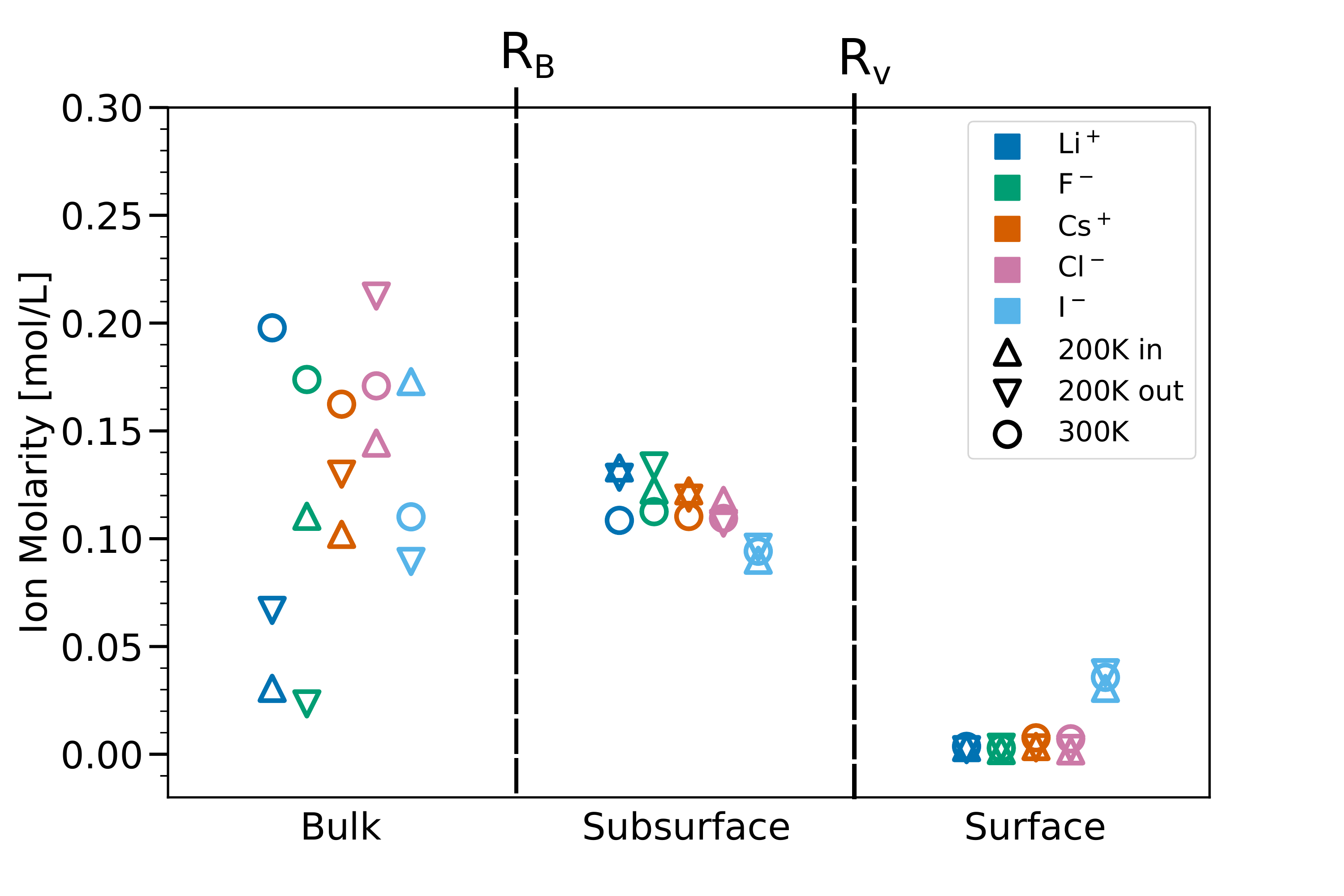}
    \caption{Ion concentrations in the core, subsurface and surface regions of an $N=776$ nanodroplet at 300~K (circles) and 200~K (triangles pointing up for simulations where the ion begins at the COM, downward for starting at the surface).}
    \label{fig:Summary-Distributions}
\end{figure}

In Fig.~\ref{fig:Summary-Distributions}, we summarize how the concentration changes in the bulk 
core ($0<r<R_B$), subsurface ($R_B<r<R_V$) and surface ($R_V<r<R_e$) regions on 
changing $T$ from 300~K to 200~K for ions other than \ce{Na+} in an $N=776$ nanodroplet. 
The details of the concentration profiles as a function of the distance from the 
droplet's COM and the histograms of the raw data 
are shown in Figs.~S6 and S7, respectively, in SI.
Fig.~\ref{fig:Summary-Distributions} shows a significant decrease in the core concentration, and  commensurate 
increase in the subsurface concentration, for \ce{Li+} and \ce{F-}, and, to a lesser extent, 
for \ce{Cs+} as well.  The larger anions \ce{Cl-} and \ce{I-} show no significant change.

In order to understand better the concentration changes at 200~K we discuss the  
direct monitoring of the transition events between the core and subsurface at 200~K, which are shown in
Figs.~S8, S9, and 10 in SI.  \ce{Li+}
and \ce{F-} show only 1-2 events entering the core over a 2~$\mu$s (considering two MD trajectories
of 1~$\mu$s each) production run, \ce{Cs+}
shows 3-4 events and \ce{Cl-}, \ce{I-} 5-6 events. During these rare events all the ions that we 
study are trapped within the core for at least
20~ns. Thus, differently from 300~K, where the large 
probability density in the core appears due to the frequent transitions 
between core and subsurface (Fig.~S11 in SI),
at 200~K, an increase in the probability density in the core appears 
due to the long residence time in
this region.

The time evolution of the \ce{Cl-} transition events 
(Figs.~S9~(a) in SI)
does not show specific preference for the interior
or the surface at both temperatures.
A similar propensity for \ce{Cl-} ions was pointed out by 
Zhao et al.\cite{zhao2013first} who performed simulations of halogen anions in a cluster of 
124 \ce{H2O} molecules at room temperature and at supercooling 
using Born-Oppenheimer (BO) MD simulations. In BO-MD the interactions are
described by quantum density functional theory (DFT). The agreement between the results supports the
validity of TIP4P/2005 to predict the location of the ions.
The direct monitoring of the transitions, Fig.~S9 in SI, shows that \ce{Cl-} and \ce{I-}
interact more than the other ions with the boundary
of the bulk core (4.0~{\AA}$\,<r<$ 7.4~{\AA}) when they are in the subsurface, 
which may be the reason that allows them to make more frequent transition into the core than other ions.  
The \ce{Cs+} ion, even though a chaotropic ion like \ce{Cl-} and \ce{I-}, shows some features
similar to \ce{Li+}, in that it does not frequently access the boundary of the bulk-like core as
\ce{Cl-} and \ce{I-} do. In the rare transitions to the core, \ce{Cs+} remains trapped for a significant amount
of time and for this reason its concentration profile appears bi-modal.
A general observation from all the trajectories is that the crossing of the
core is relatively faster than the residence time within the core, which indicates the 
presence of high potential energy barriers.  
The transition paths  can be step-wise or rapid, which reflect 
the 
variety of pathways that can be followed by the ions in and out of the core region.

\begin{figure}[htb!]
\centering
\begin{subfigure}[htbp]{0.5\textwidth}
\centering
\includegraphics[width=\textwidth]{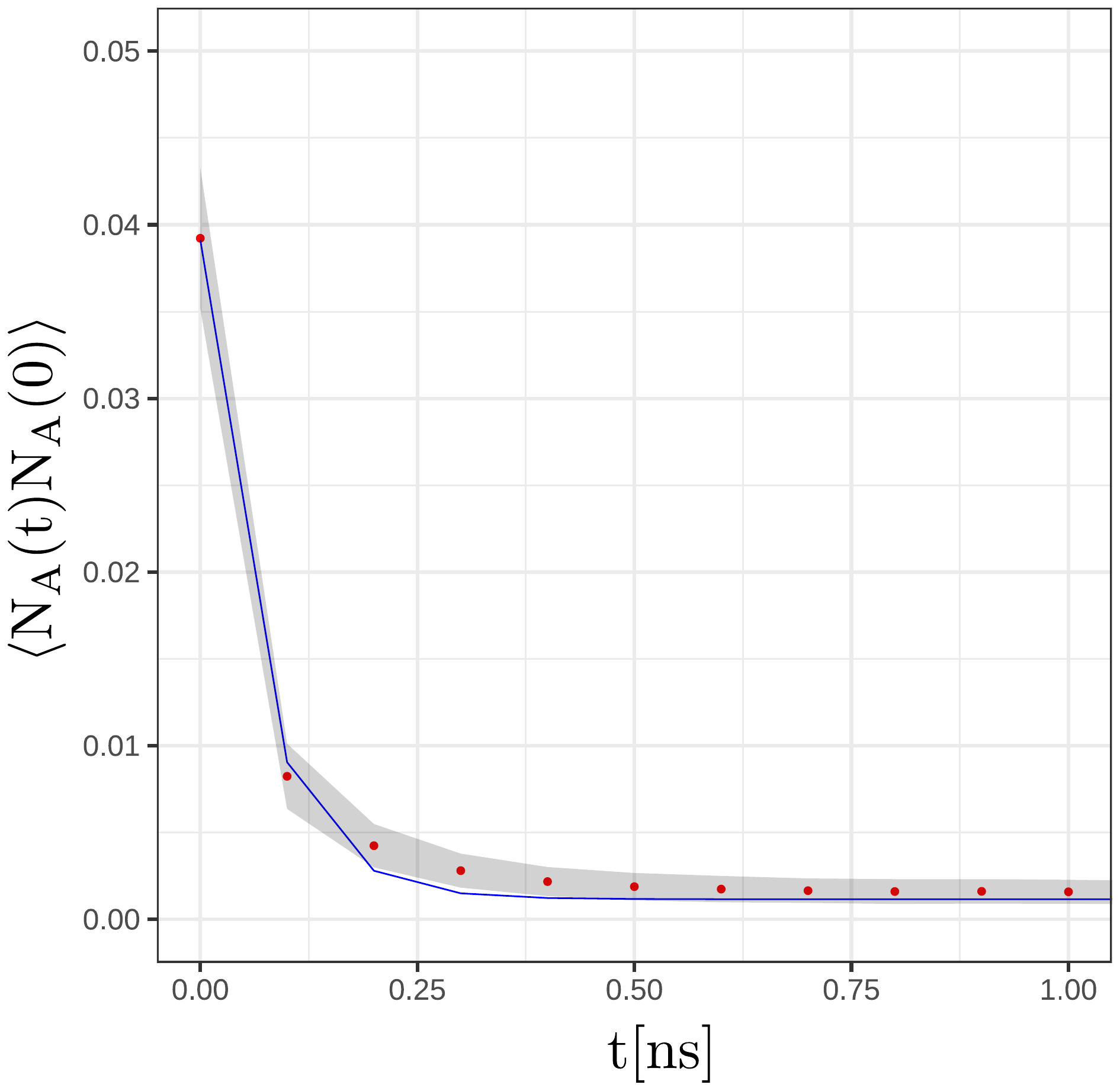}
\caption{}
\end{subfigure}
\begin{subfigure}[htbp]{0.5\textwidth}
\centering
\includegraphics[width=\textwidth]{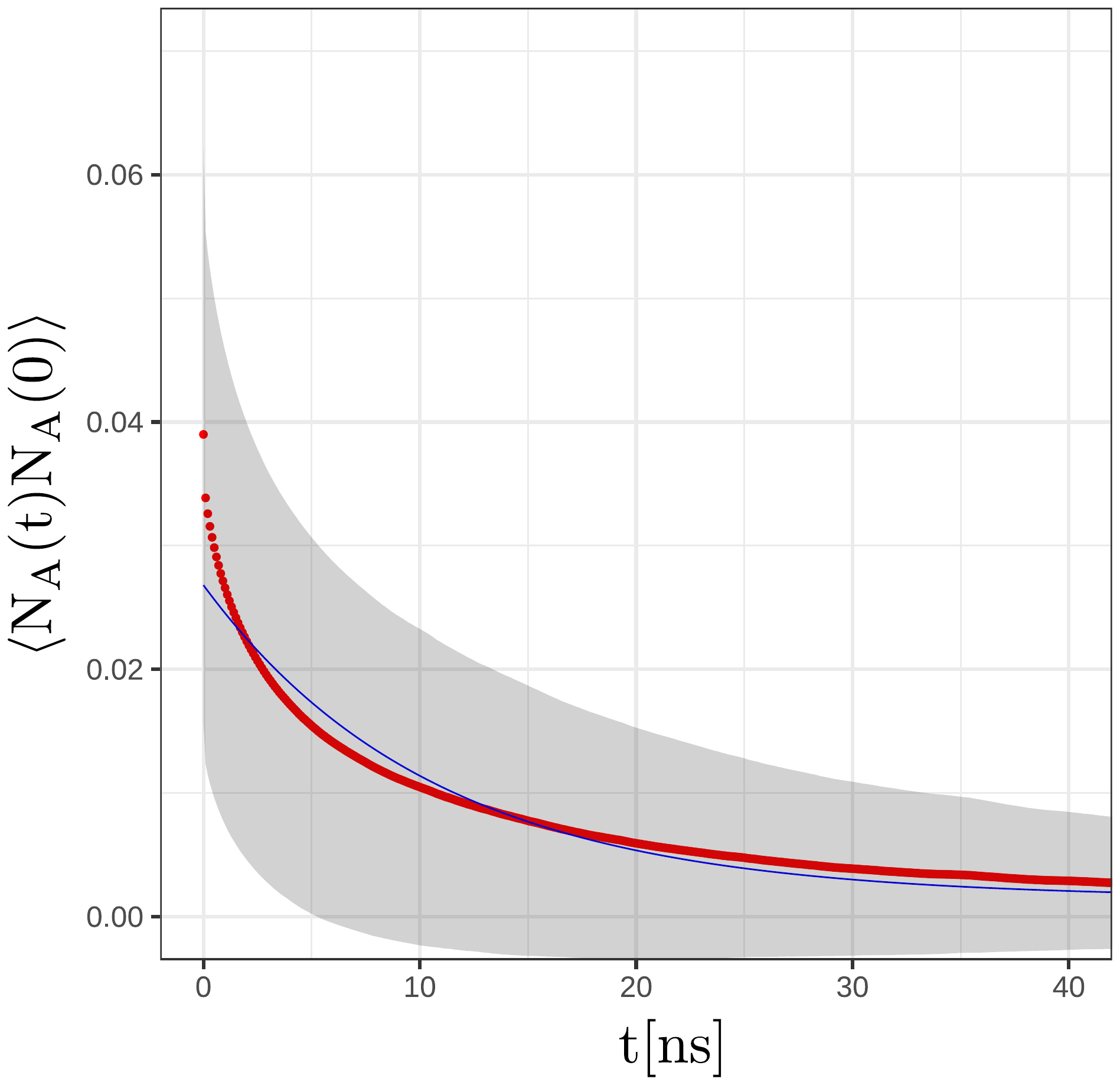}
\caption{}
\end{subfigure}
    \caption{Time correlation function (red line) of \ce{H2O} molecules coordinating \ce{F-} in $N=776$ cluster at (a) $T=300$~K and (b) $T=200$~K.
    The grey region is one standard deviation. The blue line is the exponential fitting 
    to the simulation data given by
    (a) $y = 1.76\times 10^{-3}+0.038\exp(-t/0.0637)$, (b) $y = 1.49 \times 10^{-3}+0.0253\exp(-t/10.7)$.
    }
    \label{fig:correl-flu}
\end{figure}

Now we examine the time correlation function for the exchange of the
\ce{H2O} molecules surrounding \ce{F-} in $N=776$. Following the approach of 
D.~Chandler in Ref.\cite{statistical-mechanics} we define a state
variable, $N_A$ which is 1 if a \ce{H2O} molecule is within a sphere of radius 
6.0~{\AA} with center the  \ce{F-} ion and
zero otherwise. Using this state variable we compute the correlation function $\langle N_A(t)N_A(0) \rangle$. 
The correlation functions at 300~K and 200~K are shown in Fig.~\ref{fig:correl-flu}. 
The correlation functions decay to the ratio of the number of \ce{H2O} molecules contained in
the volume of radius 6.0~{\AA} over the total number of molecules in the droplet.
We find that the decay time at 300~K is 64~ps and at 200~K 10.7~ns. The ratio of the decay time at 200~K to 300~K is
$\approx 200$ times larger than the ratio of the average velocities of the molecules at the two temperatures. 
The long residence time of the ions in the core and subsurface region is consistent with
the decay time of the correlation function.
Thus, in addition to structural and energetic factors that determine ion location, 
the two-orders-of-magnitude slowing down in dynamics at low $T$ may be necessary to 
consider when time scales relevant to an experiment become comparable to the decay time.  
Further complexities arise from the fact that dynamics at the surface and 
subsurface should be significantly faster than in the core.  
We have not addressed the radial dependence of the dynamics here, but we expect to address this in the future.

To test the effect of the force field in the different density regions,
simulations were performed with a Drude oscillator-based
force field\cite{Lamoureux2006}, which uses the SWM4-NDP \ce{H2O} model. 
The ion distribution profiles for a single \ce{Na+} and \ce{Li+} in an N=880 \ce{H2O} droplet
at 350~K and 200~K are presented in Fig.~S14 and Fig.~S15 in SI.
At high temperature, similar to the runs with TIP4P/2005, the \ce{Na+} and \ce{Li+} radial distributions 
are uniform in the core and in part of the subsurface. The \ce{Li+} shows a preference toward the
center.
In both the TIP4P/2005 and SWM4-NDP the concentration of the ions starts to increase at
1.6 nm, which indicates that the depth from the
surface is in good agreement between the two models. 
At low temperature it has been found that POL3 and SWM4-NDP  are not  suitable for ice-liquid simulation - 
they lead to a poor representation of ice\cite{gladich2012comparison, muchova2011ice}. 
We note here that the temperature in different molecular models may not correspond
to the same physical state of the cluster. 
At 200~K, simulations of bulk solution have shown that SWM4-NDP is close to the point of maximum density of water\cite{gladich2012comparison, muchova2011ice}. We have
confirmed that even in clusters at 200~K the density is at its maximum.
At 350~K, the density in the core is $32.79\pm 0.15$~$\rm{nm}^{-3}$ and in the subsurface
$32.39\pm 0.15$~$\mathrm{nm}^{-3}$ while
at 200~K, the density is $36.15\pm 0.15$~$\rm{nm}^{-3}$ and 
$35.85 \pm 0.15$~$\mathrm{nm}^{-3}$, respectively.
Interestingly, both the
\ce{Li+} and \ce{Na+} distributions using the polarizable model show their maximum in the
subsurface, similarly to the TIP4P/2005. Disregarding the unrealistic nature of the density in the cluster for SWM4-NDP, 
this example demonstrates that there may be an optimal density for solvation of the ions.

Regarding the relation of the subsurface structure in a droplet
to that of a planar surface we note that densification at supercooling also occurs for a planar interface\cite{vrbka2006homogeneous, haji2017computational}. 
In the temperature range of our study, the clusters are still in the liquid state.
The experiments of Pradzynski et al. have shown the possibility of ice formation in clusters composed of several hundreds of water molecules\cite{pradzynski2012fully}. The authors estimated that the temperature of the clusters is 90~K-115~K. We expect that in order to see a phase transition in the systems that we study, we should decreases the temperature below 200~K. In the
analysis of surface layers it 
will be challenging to distinguish small-particle surface melting from size-dependent melting\cite{peters1997surface}. 

The expulsion of the ions from the core is reminiscent of brine rejection from ice, where upon crystallization ions are expelled into the liquid~\cite{tsironi2020brine, vrbka2005brine}.  While liquid water lacks the long-range order of the crystal, the low-density form of the liquid, like ice, possesses a network structure, characterized by a first sharp diffraction peak in the structure factor arising from a high degree of tetrahedrality~\cite{elliot1991,Shieaav3194,saika2013}. 
The value of $q_T$ of deeply supercooled liquid water is comparable to that of ice~\cite{tsironi2020brine, malek2019}. For nanodroplets, the more disordered and denser subsurface appears to play the role of liquid water in usual brine rejection at freezing. We expect that the dynamics and the precise steps of the ion rejection mechanisms from the low-density core in nanodroplets will be different from those in bulk ice. It is noted that in simulations of brine rejection~\cite{tsironi2020brine}, \ce{Na+} is rejected faster than \ce{Cl-}, with \ce{Cl-} being more than twice as likely to be incorporated into the crystal lattice, an observation not dissimilar to what we see in terms of \ce{Cl-} remaining in the low-density liquid nanodroplet core, although sampling equilibrium distributions in the case of brine rejection is certainly more difficult.

\subsection{Multiple \ce{Na+} ions}

\begin{figure}[htb!]
    \centering
    \includegraphics[width=\columnwidth]{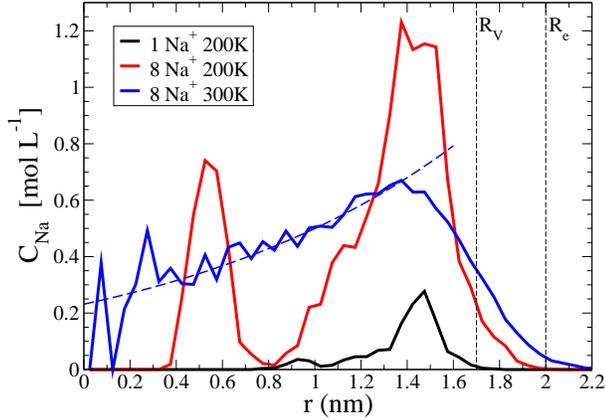}
    \caption{$C_\mathrm{Na}$ profile along the droplet radius (measured from the COM) of a single ion (at 200~K) and 8 \ce{Na+} ions (at 200~K and 300~K) in a droplet comprised 1100~\ce{H2O} molecules.  Dashed blue line is the prediction of Eq.~\ref{eq:2} for the droplet at 300~K (details in the text).}
    \label{fig:N1100}
\end{figure}

Figure~\ref{fig:N1100} shows the radial distributions of multiple ions at $T = 300$~K and 200~K in droplets comprising 1100 \ce{H2O} molecules. 
The same plot for a 
system of 776 \ce{H2O} molecules - 5 \ce{Na+} ions is shown in Fig.~S16 in SI. At $T = 300$~K the 
distributions (for 1100 and 776 \ce{H2O} molecules) are almost 
uniform with an incipient broad maximum appearing in the outer layers of the droplet.

Solution of the Non-linear Poisson-Boltzmann (NPB) equation\cite{malevanets13variation} for a 
rigid spherical geometry 
suggests that toward the droplet interior the ion distribution will show an exponential
decay
\begin{equation}\label{eq:2}
n(r) = n(R) \exp[{(r-R)/\lambda_{\mathrm{PB}}}]
\end{equation}
where $n$ is the ion number density, $R$ is the sphere radius (here taken to 
be equal to $R_e$) and $r$ is the
distance from the droplet center. $\lambda_{\mathrm{PB}}$ is given by 
\begin{equation}
\label{debye-length}
\lambda_{\mathrm{PB}} \approx \frac{\epsilon k_B T}{\sigma q} 
\end{equation}
where
$k_B$ is Boltzmann constant, $T$ is temperature,
$\epsilon$ is the permittivity of the medium, $q = m e$ is the charge of an ion
and $\sigma$ is the surface charge density given
by $\sigma = \frac{|Z|e}{4\pi R^2}$ ($|Z|e$ is the total droplet charge). 
In finding the surface charge density
we assume that all the charge is in the surface. 
The larger the $\lambda_{\mathrm{PB}}$ the slower the ion distribution decay.
One of the reasons that the simulated ion decay may
deviate from the theoretical prediction is 
the droplet's shape fluctuations. 
For this reason, we expect that the exponential decay will manifest more clearly in larger droplets at lower temperature. We note that in the larger droplets the effect of the counterions in $\lambda_{\mathrm{PB}}$
has also to be considered\cite{malevanets13variation}.
Evidently, the higher temperature will lead to a slower ion decay.
For 1100 \ce{H2O} - 8 \ce{Na+} ions at $T = 300$~K the distribution
decays (toward the droplet's COM) as an exponential function fitted
by $0.40 \exp(-(1.37-r)/1.3)$, where $\lambda_{\mathrm{PB}} \approx 1.3$~nm
(note: to convert from $n(r)$ in nm$^{-3}$ to $C_{\rm Na}(r)$ in mol/L, one should multiply $n(r)$ by 10/6.022).
Equation~\ref{debye-length} with dielectric constant of water equal to 75, yields
$\lambda_{\mathrm{PB}} \approx 0.8$. In droplets of up to a few thousands 
of water molecules the effect of shape fluctuations is significant, 
therefore, we interpret the value of $\lambda_{\mathrm{PB}}$ only in a qualitative manner. 
A value of $\lambda_{\mathrm{PB}}$ in the range of 0.8~nm (theoretical estimate)-1.3~nm 
(from fitting) is comparable in magnitude with the droplet radius ($R_e = 2.0$~nm), 
which indicates that the radial distribution function will decay slowly toward the COM. 
The predicted slow decay is indeed found in the simulations. 
The ion distribution at $T = 200$~K decays in a way that cannot be analyzed using 
the NPB predictions at the lower temperature.
The multiple ions similarly to the single ion are expelled from the drop's core 
and their distribution shows a maximum at the same location as the single 
ion (Fig.~\ref{fig:N1100}). 
The distribution shows two peaks at a distance 0.52~nm and
at 1.5~nm. The lower intensity peak at 0.52~nm corresponds to a single ion
that exchanges location with the outer ions  (found at $\approx 1.5$~nm). 
The mobility of the inner ion indicates that the appearance of the two peaks is not due to 
a metastable state.

\section{Conclusions}

We found that in supercooled aqueous droplets, a heterogeneous solvent structure leads
to a different ion radial distribution relative to that at a room temperature. 
Specifically, we demonstrated that the interior tetrahedral network that forms at supercooling expels cosmotropic ions 
(\ce{Na+}, \ce{Li+}, \ce{F-}) from the core region to the more
disordered subsurface. The radial distribution of chaotropic ions (\ce{Cl-}, \ce{I-}, \ce{Cs+}) appear to be
affected less by the presence of the bulk-structured core. They also spend most of the
time in the  subsurface from where they make rare incursions
to the core region.  Once inside the core, they reside there for a significant amount of time.

The atomistically simulated ion radial distributions are
supported by a reference analytical model that predicts the
location of an ion in a fluctuating droplet.  
The theory finds that an ion (regardless of its nature) is always subject to a harmonic potential centered at the droplet's center of mass.
This electrostatic confinement effect is more evident in certain droplet dimension and 
ion charge.  In the supercooled nanodroplets the model predictions deviate from the simulated ion distributions
due to the specific organization of the solvent that is currently not included in the analytic model.

We expect the effect of the heterogeneity in the solvent structure to appear even in the larger
droplet sizes. In microdroplets, where thermodynamics of the phase transitions
will be similar to that in the bulk solution, ice may form in the interior and then the ions
will be expelled most likely in a similar manner to the brine rejection from bulk ice.
In a planar surface, where liquid layers cover ice, the dependence of the 
thickness and ordering of the liquid surface layer on temperature has been studied
for several decades\cite{wei2001surface, sanchez2017experimental}. A similar investigation 
for the subsurface and surface layers is still missing for the nanodroplets.

The present study opens up the discussion on several questions.
Evaporative cooling of droplets is a process that has been reported
to be important in a droplet's lifetime in ionization
methods used in native mass spectrometry.
The implications of the distribution of multiple ions in the subsurface in 
a Rayleigh fission and ion-evaporation mechanism have to be examined.
We have found that a conical deformation is a key structure
in the mechanism of ion emission\cite{kwan2021relation}.
An intriguing question is whether
the formation of a conical deformation in the droplet that emits ions 
is affected by the low density liquid at supercooling or an ice interior. 
A related question is whether the Rayleigh limit of a supercooled droplet
changes since the conducting region for certain ions
is restricted to the subsurface and surface instead of exploring the volume of the entire droplet.
In a long time, the droplet is expected to be a conductor, since the
ion still make rare visits to the core.
In an intermediate length of time, which may be critical for charging of
macromolecules of instance, the properties of the droplet may deviate from those of 
a good conductor.
A study on these questions will be relevant
to atmospheric chemistry, native mass spectrometry and the physics of jets.

We anticipate our study to provide the starting point for investigating the
structure of the supercooled droplets containing unstructured peptides 
and complexes of ions. Several possible scenarios
are envisioned that may affect the charge states of 
macromolecules. On the one hand, if a macromolecule and simple ions are expelled to the subsurface,
the macromolecule will be exposed to 
a higher concentration of ions and different structure of solvent than that at room or elevated temperature. 
Thus, their charge state and release mechanism will be determined
by the distinct chemical features of the subsurface. 
Because of the higher concentration of
ions in the subsurface than at room or elevated temperature, 
for certain macromolecules, a ``salting-out'' release pathway may
open up that expels them to the vapor phase. Clusters containing solvated 
NaCl or \ce{[Na_2Cl]+} complexes where the ions
are in their contact-ion pair form may be 
also expelled. This is a new idea that has not been explored before
as a possible mechanism of release of macroions from droplets. If macroions are released
via a salting-out path, or if the counterions are released in the form
of complexes with the co-ions they will not form adducts with the macromolecules. 
This outcome is in
agreement with still unresolved observations in native mass spectrometry experiments.
On the other hand, if a macromolecule (regardless of its shape)
is trapped in a glassy core
its diffusion toward
the surface will be delayed and its exposure to ions will be altered, which will in turn 
affect its charge state. 
In this case, a charged-residue mechanism is more likely to be followed in 
the charging of the macromolecule.
We currently explore these new possible macroion release mechanisms.
We envisage that studies in this direction will assist in interpreting ion 
mobility-mass spectrometry data on the detection of the conformations
of macromolecules originating from the bulk solution and will provide insight 
into the chemistry of atmospheric aerosols.

\section*{Supporting Information} 

(S1) Derivation of the electrostatic confinement (EC) model and comparison with atomistic simulations of an ion in a droplet. 
(S2) Details of the computational methods and models. 
(S3) Water density and structure, and single \ce{Na+} radial concentration for various droplet sizes 
(S4) 
Evidence of the
convergence of the simulation data, and 
radial probability density (concentration) profiles 
for \ce{I-}, \ce{Cl-}, \ce{F-}, \ce{Li+}, \ce{Cs+} in room temperature and supercooled nanodroplets. 
(S5) Radial distribution functions for ion-Oxygen of \ce{H2O} and ion-Hydrogen of \ce{H2O}. 
(S6) Radial probability density (concentration) profiles of ions in supercooled and room temperature 
nanodroplets using a polarizable molecular model.
(S7) Multiple-ion radial concentration profiles in supercooled and room temperature nanodroplets.

\section*{Acknowledgments}
S.C. thanks Prof.~D.~Frenkel, Department of Chemistry, University of
Cambridge, UK, and
Dr. Anatoly Malevanets, The University of Western Ontario for discussions on
the stability of charged systems. Professor D. Russell, Department of Chemistry,
Texas A \& M University is thanked for pointing out the variable temperature ESI source and the
role of cold droplets in transferring the protein conformations for MS analysis. 
V.K. acknowledges the province of Ontario and the
University of Western Ontario for the Queen Elizabeth II Graduate Scholarship
in Science and Technology.
I.S.-V.~thanks the Departments of Applied Mathematics and Chemistry at Western University for sabbatical hosting.  
We acknowledge the financial support from Natural Sciences and Engineering Research Council (Canada).  Computational resources were provided by ACENET and Compute Canada.

\newpage

\providecommand{\latin}[1]{#1}
\makeatletter
\providecommand{\doi}
  {\begingroup\let\do\@makeother\dospecials
  \catcode`\{=1 \catcode`\}=2 \doi@aux}
\providecommand{\doi@aux}[1]{\endgroup\texttt{#1}}
\makeatother
\providecommand*\mcitethebibliography{\thebibliography}
\csname @ifundefined\endcsname{endmcitethebibliography}
  {\let\endmcitethebibliography\endthebibliography}{}

\clearpage


\end{document}



\section{S1. Details of the analytical model of a single ion location within a fluctuating droplet}
Here we discuss details of the analytical model that predicts a simple ion's or macroion's location in
a fluctuating droplet. The key points of the model are presented in the main text.

The following discussion is an extension of our previous work on the energy of a
continuum dielectric droplet containing a single 
(macro)ion\cite{consta2015disintegration, oh2017droplets}.
In the model the droplet surface fluctuations are considered.
The total energy of the droplet ($E$) is written
as the sum of surface energy ($E_{\mathrm{surf}}$) and electrostatic energy ($E_{\mathrm{el}}$)\cite{rayleigh1882, oh2017droplets},
\begin{equation}
  \label{eq:energy}
  E = E_{\mathrm{surf}} + E_{\mathrm{el}} = \gamma A + E_{\mathrm{el}}
\end{equation}
where $\gamma$ is the surface tension and $A$ surface area. $E_{\mathrm{el}}$ is
given by
\begin{equation}
  E_{\mathrm{el}} = -\frac{1}{2} \int\limits_{\mathbb{R}^3/V} d\mathbf{r} 
  (\epsilon^{E}-\epsilon^{I}) \mathbf{E}\cdot\mathbf{E}_0
  \label{eq:energy-electr}
\end{equation}
where $\mathbb{R}^3/V$ are the points in the exterior of the droplet, $\epsilon^{I}$ is the electric permittivity in the interior of a droplet, $\epsilon^{E}$ is the electric permittivity
of the medium surrounding the droplet, and
\begin{equation}
  \mathbf{E}_0(\mathbf{r}) = -\nabla \frac{Q}{4\pi\epsilon^{I}r} .
  \label{eq:e0}
\end{equation}

The distance of a point on the droplet surface from the ion is given by:
\begin{equation}
  \label{eq:r-polar}
  \rho(\sigma) = R+\sum_{l>0,m_l}a_{l,m_l}Y_{l,m_l}(\sigma)
\end{equation}
where $\sigma = (\theta,\phi)$ is the spherical angle, $\rho(\theta,\phi)$ is the
distance from the centre (which is at the ion), and $Y_{lm}(\theta,\phi)$ denote the
spherical harmonics functions of rank $m$ and order $l$. For 
certain shapes of droplets, such as bottle-necked shapes or shapes like
an eight we should choose the center of the shape carefully, 
so as we do not have for a single $(\theta,\phi)$ more than
one values of $\rho$. In other words, the same line intersects the
shape in several points.
$R$ is the $l =0$ term in the expansion of $\rho(\sigma)$.
The details of the algebra for expressing $E_{\mathrm{surf}}$ in terms of the
expansion coefficients $a_{l,m_l}$ (see Eq.~\ref{eq:r-polar}) 
is given in Ref.[\cite{consta2015disintegration}]. The coupling of the
electrostatic energy\cite{jackson} to the shape fluctuations is a tedious step and one of the
ways to do that is found in Ref.[\cite{oh2017droplets}].

After some algebra, the total energy is given by
  \begin{equation}
    \label{eq:en-total}
\begin{aligned}
    E = & \frac{(\epsilon^{I}-\epsilon^{E})Q^2}{8\pi\epsilon^{I}
         \epsilon^{E}R_0} 
        \left[ 1 - \sum_{l>0,m_l} 
       \frac{\epsilon^{I}l(l-1) - \epsilon^{E}(l+1)(l+2)}
      {\epsilon^{I}l + \epsilon^{E}(l+1)} \frac{|a_{l,m_l}|^2}{4\pi R_0^2} \right] \\ 
      &   + \gamma \left [ 4 \pi R_0^2 + \frac{1}{2} \sum_{l>0, m_l} (l - 1)
      (l+2) |a_{l, m_l}|^2 \right ].
\end{aligned}
  \end{equation}

We will show that the $l= 1$ term
in  Eq.~\ref{eq:en-total} depends on the distance squared of the ion from the droplet COM. 
In the algebra that follows we will use that
\begin{equation}
  \label{eq:r-fourth}
  \rho^4(\sigma) = R^4+4R^3\sum_{l>0,m_l}a_{l,m_l}Y_{l,m_l}(\sigma) + \cdots .
\end{equation}
In Eq.~\ref{eq:r-fourth} we keep only the two dominant terms in the summation. The remaining of the terms  
are neglected because they include powers $ \ge 2$ of $\delta r = \sum_{l>0,m_l}a_{l,m_l}Y_{l,m_l}(\sigma)$
($\delta r$ is a small perturbation relative to $R$).

We find the coordinates $X_{\mathrm{COM}}, Y_{\mathrm{COM}}, Z_{\mathrm{COM}}$ of the 
droplet's COM in
terms of the expansion coefficients $a_{l,m_l}$. In the following expressions 
$d\sigma = \sin\theta d\theta d \phi$.
\begin{equation}
\label{eq:com}
\begin{aligned}
Z_{\mathrm{COM}}={} & \frac{1}{V} \int Z(r, \theta, \phi) d^3 r =  \\
      &  \frac{1}{V} \int_{r \le \rho, \sigma \in S^2} r \cos \theta r^2  d\sigma dr = \\
      &   \frac{1}{4V} \int \rho^4 (\theta, \phi) \cos \theta   d\sigma = \\
      &  \frac{1}{V} R^3 \int \cos \theta \sum_{l>0,m_l}a_{l,m_l}Y_{l,m_l}(\sigma)  d\sigma = \\
      & \left (\frac{3}{4\pi} \right )^{1/2} a_{1,0}
\end{aligned}
\end{equation}
In the fourth line of Eq.~\ref{eq:com} we use the orthogonality of the spherical harmonics.
Similarly, $X_{\mathrm{COM}} = \Re(a_{1,1})\sqrt2$ and $Y_{\mathrm{COM}} = \Im(a_{1,1})\sqrt2$.

The $l = 1$ term in Eq.~\ref{eq:en-total} yields
\begin{equation}
\label{eq:energy-l1}
\Delta E_1 = \frac{(\epsilon^{I}-\epsilon^{E})Q^2}{8\pi\epsilon^{I}
      \epsilon^{E}R_0} 
      \frac{6\epsilon^{E}}
      {\epsilon^{I} + 2\epsilon^{E}} \frac{(|a_{1,0}|^2+|a_{1,1}|^2+|a_{1,-1}|^2)}{4\pi R_0^2} .
\end{equation}
Using Eq.~\ref{eq:com} and the similar ones for $X_{\mathrm{COM}}$ and $Y_{\mathrm{COM}}$, Eq.~\ref{eq:energy-l1} becomes
\begin{equation}
\Delta E_1(\|\mathbf{r}\|) =
  \frac{\varepsilon-1}{4\pi\epsilon_0\varepsilon(\varepsilon+2)}	
  \frac{Q^2}{R^3}
  \|\mathbf{r}\|^2 
  \label{eq:ener-ion}
\end{equation}
where $Q$, $R$ and $\varepsilon$ are the charge of the ion, the
droplet radius and the relative dielectric constant of the solvent,
respectively, $\varepsilon_0$ is the vacuum permittivity
and $\|\mathbf{r}\|^2 = X^2_{\mathrm{COM}} + Y^2_{\mathrm{COM}}  +Z^2_{\mathrm{COM}}$.
The coefficient in front of $\|\mathbf{r}\|^2$ is denoted as $K (\varepsilon)$ and
we call it ``spring constant''. Thus,
\begin{equation}
  K(\varepsilon) = 
  \frac{\varepsilon-1}{4\pi\epsilon_0\varepsilon(\varepsilon+2)}        
  \frac{Q^2}{R^3} .
  \label{eq:Ke} 
\end{equation}
The plot of the variation of $K(\varepsilon)$ as a function of $\varepsilon$ 
is found in the main text.

\begin{figure}
  \begin{subfigure}[htbp]{0.45\textwidth}
        \centering
                \includegraphics[width=\linewidth]{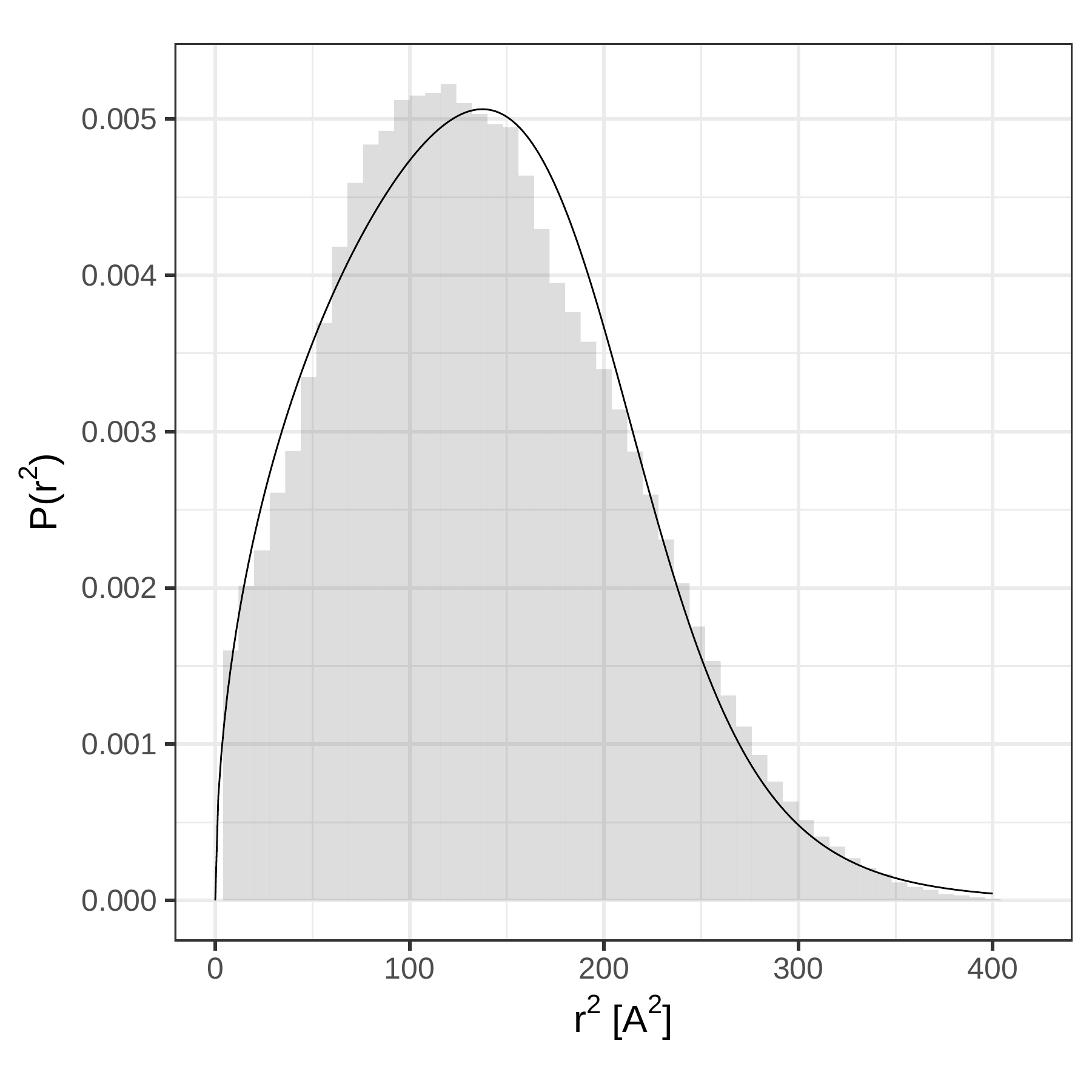}
                \caption{}
        \end{subfigure}

       \begin{subfigure}[htbp]{0.45\textwidth}
                \centering
                \includegraphics[width=\textwidth]{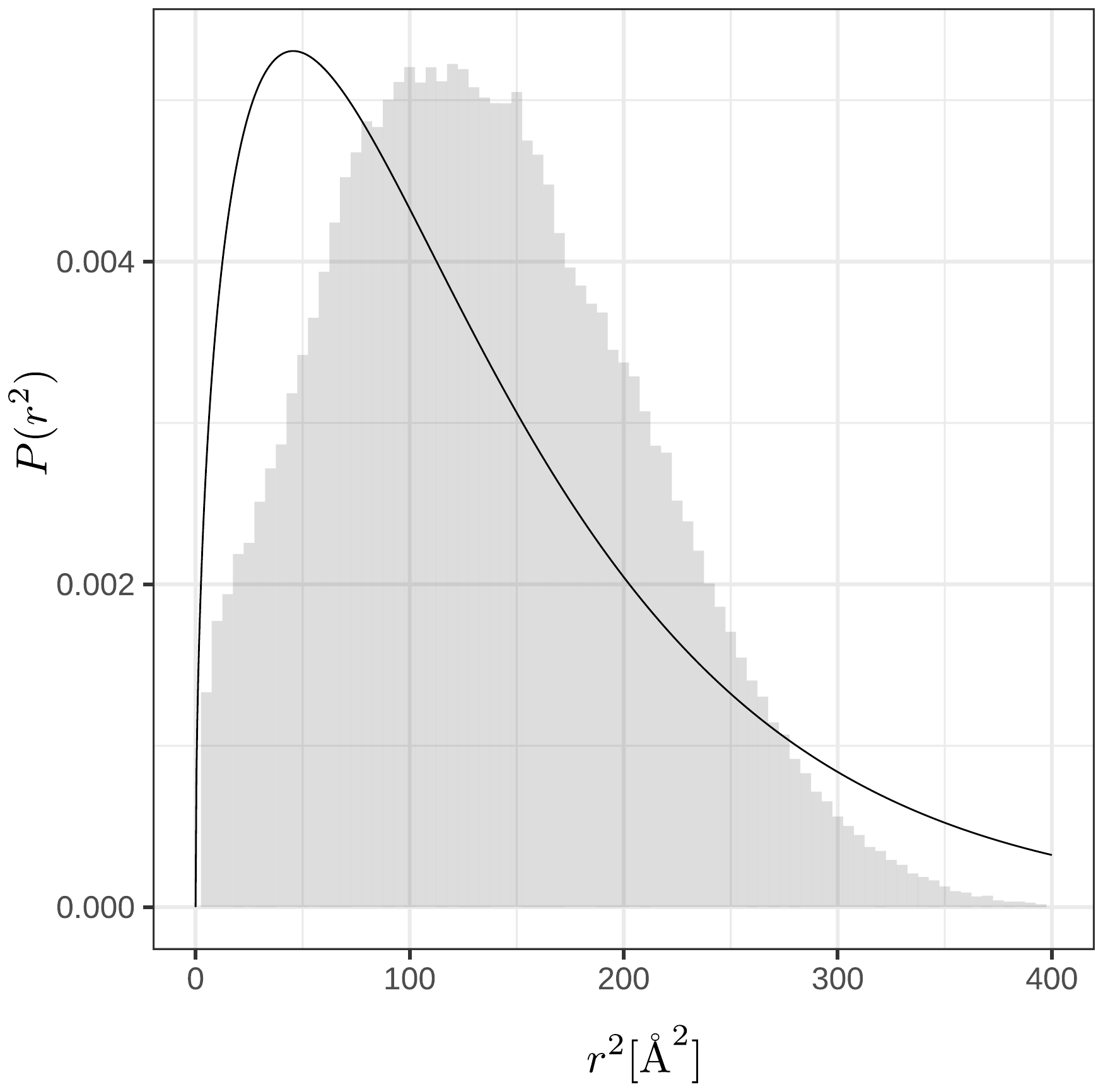}
                \caption{}
        \end{subfigure}
  \caption{(a) Distribution of the distances of the sodium ion
    relative to the COM of the droplet. The equimolar droplet radius
    is 1.9~nm. The solid line is the product of the logistic and the
    gamma distribution (with shape parameter 1/2) intended to capture confinement effects. The
    function is fitted to the distribution using the maximum likelihood estimate (MLE). 
    (b) Same as (a) but
    the solid line is the gamma function fitted to the
    distribution using MLE.}
  \label{fig:sod}
\end{figure}

If the ion is localized in the center of a droplet the Gibbs-Boltzmann
distribution of the ion positions is given by
\begin{equation}
  P(\|\mathbf{r}\|^2) = \frac{2}{\sqrt{\pi}}
  \left(\frac{K(\varepsilon)}{k_{B}T}\right)^{3/2} \|\mathbf{r}\|
  e^{-K(\varepsilon)\|\mathbf{r}\|^2/k_{B}T} . 
  \label{eq:gibbs}
\end{equation}
The expectation value of the square of the distance of the ion from
the droplet's COM is given by
\begin{equation}
  \langle \|\mathbf{r}\|^2 \rangle = \frac{3}{2} \frac{k_{B}T}{K(\varepsilon)}.
  \label{eq:r2-exp}
\end{equation}
The conditions under which the EC is more pronounced are discussed in the main text.

We assume that the number density of the solvent in the vicinity of
the droplet surface is well approximated by the logistic function
(\ref{eq:logistic})
\begin{equation}
  n_{r_0,d}(r) = \frac{1}{1+\exp(-(r-r_0)/d)} \label{eq:logistic}
\end{equation}
where $d$ and $r_0$ are fitting parameters that can be interpreted as
the droplet radius and the width of the surface layer.  Using the
logistic curve for for the number density (entropic factor) and gamma
function that takes into account the electric potential (energetic
factor) we arrive at the following ansatz for the ion distribution
\begin{equation}
  p(r^2) \sim r e^{-K r^2/k_{B}T} \frac{1}{1+\exp(-(r-r_0)/d)}
  \label{eq:total-distr}
\end{equation}

Using the maximum likelihood estimate (MLE) approach we found the most probable
parameters ${\{K,r_0,d\}}$ in order to match the observed values of
the distance of the ion from the center of mass. In
Fig.~\ref{fig:sod} we show the fitted and the observed distributions
for a single sodium ion in a droplet of 1000 TIP3P (transferable intermolecular potential with three
points)\cite{jorgensen1998temperature} water
molecules. The data were obtained in 15~ns molecular dynamics
simulations using the NAMD package\cite{phillips05scalable}.  The fittings were produced with
the use of statistical analysis software R\cite{Rpackage}. For comparison
we contrast the fit that takes into account the surface of the droplet
with a fit to a gamma distribution of shape $1/2$ in
Fig. \ref{fig:sod}~(b). The analysis shows that the shape
fluctuations of the droplet accounts for the distribution of the ions
in the droplet.  The fitting can only establish an upper bound of the
parameter $K$.  All the variability of the charge distribution is
explained by the confinement effect of the droplet surface. In
Fig.~\ref{fig:sod} the fitting parameters used in
Eq. (\ref{eq:total-distr}) are ${r_0=\mathrm{14.9\AA}}$ and
${d=\mathrm{1.1\AA}}$. The effective radius of the droplet is smaller
than that of the true molecular surface of water.

Here we demonstrate the effect of geometric vs electrostatic confinement in a droplet
composed of 1000 \ce{H2O} molecules and a single ion with charge 1\ce{e+} and
3\ce{e+}. The data are summarized in Table~S1.

For a droplet composed of 1000 \ce{H2O} molecules (equimolar radius = 1.93~nm) 
and a 1\ce{e+} ion, if we assume $\varepsilon = 80$ at $T=300$~K then 
Eq.~\ref{eq:Ke} yields $K = 0.39$~${\mathrm{[mJ/m^2}]}$ and 
Eq.~\ref{eq:r2-exp} yields 
${\langle \|\mathbf{r}\|^2 \rangle} = 16$~${[\mathrm{nm^2}]}$.
The simulations for the same system yield 
${\langle \|\mathbf{r}\|^2 \rangle} = 1.36$~${[\mathrm{nm^2}]}$.
These data are shown in the first line of Table~S1.
The fact that the estimated value of ${\langle \|\mathbf{r}\|^2 \rangle}$ is larger than the droplet's 
radius squared implies that the geometric confinement dominates over the electrostatic confinement.

In the second line of Table~S1, we use the ${\langle \|\mathbf{r}\|^2 \rangle} = 1.36$~${[\mathrm{nm^2}]}$ (from
simulations) and the data are fitted with a gamma distribution to yield $K$. 
In the third line of Table~S1, the same dated are fitted with Eq.~\ref{eq:total-distr}, which yields an 
upper bound for $K$.

In the fourth and fifth lines of Table~S1, we show data for a droplet composed of 1000~\ce{H2O} molecules
and a 3\ce{e+} ion. The fitting with a gamma distribution of the simulation data yields
$K = 20.7$~${ \mathrm{[mJ/m^2}]}$.
This value of $K$ yields $\varepsilon = 38$.
In the fifth line, we estimate the value of $K$ from Eq.~\ref{eq:r2-exp}, where ${\langle \|\mathbf{r}\|^2 \rangle}$
is the value estimated from the simulations. Thus, estimation of $K$ by two independent
ways yield very similar. The similarity indicates that the electrostatic confinement becomes
significant for this system.

The estimated values of the dielectric constant ($\varepsilon = 38$) is lower than the
typical values of the pure solvent. We believe that the apparent
decrease in the dielectric constant is connected with the polarization saturation
in the vicinity of the charge ion.

\begin{table*}
  \centering%
  \begin{tabular}{l|r|r|r|r}
    & Size [N]
    & Charge [Q]                           & $K$ ${ \mathrm{[mJ/m^2}]}$
    & ${\langle \|\mathbf{r}\|^2 \rangle}$
      ${[\mathrm{nm^2}] }$
    \\
    \hline
    Theor. $\epsilon=$80 & 1000
    & 1 \ce{e+}                            & 0.39                                   & 15.3
    \\
    Sim. Gamma           & 1000
    & 1 \ce{e+}                            & 4.53                                   & 1.36
    \\
    Sim. Confinement     & 1000
    & 1 \ce{e+}                            & $< \mathrm{0.83}$                      & 1.36
    \\
    Sim. Gamma           & 1000
    & 3 \ce{e+}                            & 20.7                                   & 0.30
    \\
    Theor. $\epsilon=$38 & 1000
    & 3 \ce{e+}                            & 20.5                                   & 0.30
    \\
  \end{tabular}

  \caption{Values of the parameter $K$ for selected simulations of an
    ion in a droplet. All the estimates are at $T=300$~K. Details are presented in the text.}
  \label{tab:kappa}
\end{table*}
\newpage

\begin{figure}
                \includegraphics[width=0.5\linewidth]{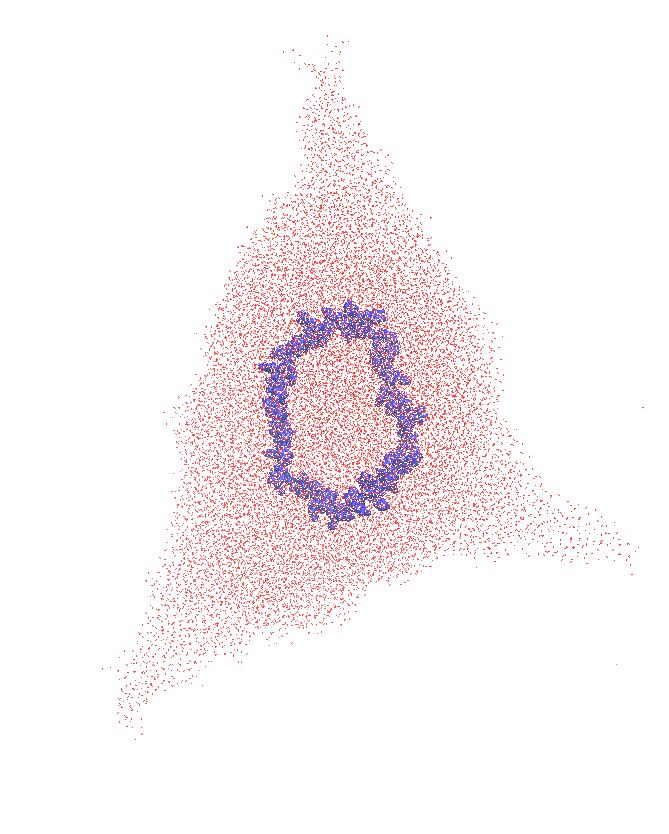}
  \caption{Snapshots of aqueous droplets with a charged cyclic peptide. 
    Spikes formed when the charge squared-to-volume ratio is 
  above a certain threshold value defined by the Rayleigh limit.}
   \label{fig:cyclic-peptide}
\end{figure}

The results of the simulations of a sodium ion in water droplet can be
compared with the results of the simulations of a cyclic peptide in Fig.~2 in the main text. 

\newpage

\section{S2. Models and simulation methods}

\begin{table*}
 \begin{tabular}{ | c | c | c | c| c | c | c | c| }
   \hline
    \multicolumn{1}{|c|}{$N$} & $L$ & $T$ & $N_{\mathrm{Na}}$  & $t_{\mathrm{run}}$ & {$N_{\rm d}$} & $R_e$ & $\tau$  \\ 
     \multicolumn{1}{|c|}{(\ce{H2O})} &  (nm) & (K) & & ($\mu$s) & & (nm)   & (ns)\\ \hline \hline
     \multirow{2}* {100} & 10 & 200 & 1 & 1.6 & 100 &  0.87 &  0.4 \\
      & & 260 & 1 &  1.6 &  99.8 &  & 0.8 \\    \hline
   \multirow{2}*{200} &  10 & 200 & 1 & 1.6 & 200 & 1.10 & 0.8  \\
      & & 260 & 1 & 1.6 & & 199.8 & 0.8   \\    \hline
   \multirow{3}*{360} &  10 & 200 & 1 &  1.6 & 360 & 1.35 & 0.4  \\
      & & 200* & 1 & 0.44 &  360  & 1.35 &0.4\\
      & & 300 & 1  & 1.6 & 359.3 & 1.35 & 0.2  \\      \hline
   \multirow{3}*{776} &  15 & 200 & 1 & 0.33 & 776 &   1.77 &  0.8 \\
      & & 200* & 1 & 0.46 & 776 & 1.77 & 0.8 \\
      & & 200* & 5 & 0.53 &  776 & 1.77 & 0.8 \\
      & & 300  & 1 & 0.77 &  773.9 & 1.75 & 0.4  \\      \hline
   \multirow{2}*{1100} & 20 & 200 & 1 & 0.32  & 1100 & 2.0  & 1.6 \\
      & & 200* & 8 &  0.055 &  1100 & 2.0  & 1.6 \\
      & & 300 & 1  &  0.32  &  1095.1 & 1.97  & 0.8  \\      \hline
 \end{tabular}
\caption{Simulation parameters. $N$ denotes the number
of \ce{H2O} molecules in the simulation box of dimension $L$. 
$N_d$ is the average number of the \ce{H2O} molecules that form
a connected drop, $\tau$ is an upper bound on the relaxation time and $t_{\mathrm{run}}$ is the 
duration of the run. The ``*'' superscript in the temperature refers to simulations started with \ce{Na+} ion(s) in or near the droplet center. $R_e$ denotes the equimolar radius.}
 \label{simucond1}
 \end{table*}

\begin{table*} 
 \begin{tabular}{|c|c|c|c|}
	\hline
Ion                                  & Charge ($e$)    & $\epsilon_{\mathrm{LJ}}$ (kJ/mol) & $\sigma_{\mathrm{LJ}}$ (nm) \\
	\hline
\ce{Na+} (Ref. \cite{aqvist1990ion})         &  $+1$   &  0.0115980     & 0.333045  \\
\ce{Li+} (Ref. \cite{aqvist1990ion})         &  $+1$   &  0.0764793     & 0.212645  \\
\ce{Cs+} (Ref. \cite{aqvist1990ion})         &  $+1$   &  0.000338904   & 0.671600  \\
\ce{F-}  (Ref. \cite{Chandrasekhar1984})     &  $-1$   &  3.01248       & 0.273295  \\ 
\ce{Cl-} (Ref. \cite{Chandrasekhar1984})     &  $-1$   &  0.492833      & 0.441724  \\
\ce{I-}  (Ref. \cite{McDonald1998})          &  $-1$   &  0.292880      & 0.540000  \\
	\hline
\end{tabular} 
\caption{Charge and Lennard-Jones (LJ) parameters ($\sigma_{\mathrm{LJ}}$ representing the atomic diameter and $\epsilon_{\mathrm{LJ}}$, depth of the potential energy minimum) for the ions used with the TIP4P/2005 water model.} 

\label{table:LJ-param}
\end{table*}

\paragraph{A. Simulations of aqueous nanodroplets with \ce{Na+} ions} We simulate \ce{Na+} ions in aqueous nanodroplets at $T=$200~K, 260~K, and 300~K, representing 
the room temperature and supercooled conditions. 
The system sizes and length of simulations are shown in Table~\ref{simucond1}.
The simulations were performed by molecular dynamics (MD) as implemented in 
 GROMACS v4.6.1~\cite{Berendsen1995,Lindahl2001,van2005,Hess2008}.
The water molecules were modeled with the TIP4P/2005 (transferable intermolecular
potential with four points) model~\cite{Abascal2005}. 
The \ce{Na+} parameters are shown in Table~\ref{table:LJ-param}. The interactions of the ion with the O site of the water molecules are calculated with the combining rules $\mathrm {\epsilon_{Na,O}=\sqrt{\epsilon_{Na}\epsilon_O}}$ and $\mathrm {\sigma_{Na,O}=\sqrt{\sigma_{Na}\sigma_O}}$.

Each nanodroplet 
has been placed in a periodic cubic box of length $L$
(see Table~S2). The box is large enough to avoid any interaction between the water droplet and its periodic images. The length of cutoff for interactions (Coulomb and Lennard-Jones) is at $L/2$, which is much larger than the droplet's diameter in order to reproduce long range electrostatic interactions within the droplet.
The temperature was controlled with the Nos{\'e}-Hoover thermostat with time constant 0.1~ps.
The equations of motion are integrated with the leap-frog algorithm with a time step of 2~fs.

The simulations were initiated with a condensed pure water nanodroplet where the \ce{Na+} ion were placed at the surface for the majority of the single \ce{Na+} runs, and  in the center for two runs, $N=360$ and 776  at $T=200$~K. All runs with multiple \ce{Na+} ions start with the ions near the droplet center of mass.  

In Table~\ref{simucond1}, the mean number of molecule, $N_d$, forming the connected cluster (i.e. those not in the vapor), and the relaxation time $\tau$, determined from the neighbor correlation function are shown\cite{maleknc2018}. The values of $\tau$ provide an estimate for the relaxation time for simulations that include ions. In the temperature range where simulations are performed the solvent evaporation within the simulation box is negligible.

To ensure that we sufficiently sample an equilibrated system after the addition of a single Na$^+$ ion at $200$~K, where the concern for equilibration is the highest, we run two  simulations for each of $N=360$ and $N=776$ nanodroplets.  In one set, we set the \ce{Na+} ion at or near the centre of the droplet, quench the system through a conjugate-gradient energy minimization, and then proceed with an MD simulation.  In the other, we initially place the ion on the surface.  Equilibration time is estimated from the time it takes for the results of the simulations from the two different conditions to converge. For example, for $N=776$ after 400~ns, the ion densities as a function of radial distance from the droplet COM $\rho_{\rm Na}(r)$ converge for the two simulations.  For $N=1100$, we assume that the equilibration time is longer by a factor of $\tau_{1100}/\tau_{776}\approx 2$.  For multiple ions, initially distributed in the nanodroplet interior, we presume that the relaxation time is shorter  and that the single ions simulations provide upper bounds on the relaxation times. 

\paragraph{B. Simulations of aqueous nanodroplets with \ce{Li+}, \ce{Cs+} and anions} We performed MD simulations of droplets comprised 100 and 776 \ce{H2O} molecules and a single \ce{F-}, \ce{Cl-}, \ce{I-}, \ce{Li+}, \ce{Cs+} ion. 
The simulations were performed with NAMD 2.14\cite{phillips05scalable}.
The water molecules were modeled with the TIP4P/2005 model~\cite{Abascal2005}.
The ion Lennard-Jones parameters are shown in Table~\ref{table:LJ-param}.

The Newton’s equation of motion for each atomic site was integrated using the 
velocity-Verlet algorithm with a time step of 2.0 fs. 
All the forces were computed directly without any cut-offs. Equilibrium simulations in NAMD were set 
by placing the droplet in a spherical cavity of radius 20.0 nm by using spherical boundary condition.
The systems were thermalized with the Langevin thermostat with the damping coefficient set to 1/ps.
The simulation included a 0.2~$\mu$s equilibration period followed by a 1.0~$\mu$s production run, 
with configurations sampled  every 0.1~ns. At 200~K two simulations started with the ion
placed initially near the COM and on the surface. 

\paragraph{C. Simulations of aqueous nanodroplets with polarizable force field}
We performed MD simulations of droplets comprised 880~\ce{H2O} molecules and a single
\ce{Na+} and \ce{Li+} ion at 350~K and 200~K.
The simulations were performed by using the software NAMD version 2.14\cite{phillips05scalable}. 
The water molecules were modeled with the SWM4-NDP model\cite{Lamoureux2006} 
and the ions were modelled with the CHARMM Drude force field \cite{Yu2010,Luo2013}. 
The SWM4-NDP model is a 5-site model with four charge sites and a negatively charged Drude particle connected to the oxygen atom, while the ions are modeled with one charge site
and a negatively charged Drude particle. Hereafter, we will refer to the SWM4-NDP model as SWM4 for brevity.
A dual Langevin thermostat was utilized to freeze the Drude oscillators while maintaining
the warm degrees of freedom at the desired temperature\cite{Jiang2011}. 
The systems were thermalized with Langevin thermostat at 350~K (for the warm degrees of freedom) 
and at 1~K for the Drude oscillators. The damping coefficient for the Langevin thermostat was 
set to 1/ps. The length of the production run was 100~ns, sampled every 0.5~ps for high temperature runs and every 100~ps for low temperature runs. 
The simulation protocol was the same as for the ions in S2.B.

\newpage

\section{S3. Water density and structure, and single \ce{Na+} radial concentration for various droplet sizes}
\label{sec:SIsizes}

\begin{table*} 
 \begin{tabular}{|c|c|c|c|}
	\hline
Ion & $N$, $T$~(K) & number density ($\mathrm{nm}^{-3}$)    & number density ($\mathrm{nm}^{-3}$)   \\
    &              &              at $0< r< R_\mathrm{B}$   & at $R_\mathrm{B} < r < R_\mathrm{V}$ \\
	\hline
\ce{Na+}  & 776, 200~K   &  33.66     &  34.56  \\
          & 776, 300~K   &  34.38    &  34.14    \\ 
          & 1100, 200~K  & 33.08     & 34.11      \\
          & 1100, 300~K  &  34.26    &  34.03  \\
\hline
\ce{F-}  & 776, 200~K   & 33.71     & 34.60  \\
	\hline
\ce{Cl-}  & 776, 200~K   & 33.37     & 34.40  \\
          & 776, 300~K   & 34.13     & 33.98 \\
\hline
\end{tabular} 
\caption{Number density based on the number of oxygen sites in the core ($0< r< R_\mathrm{B}$) and subsurface ($R_\mathrm{B} < r < R_\mathrm{V} $). The error in the densities is $\pm 0.15$.}
\label{table:densities}
\end{table*}


\begin{figure}
    \centering 
     \includegraphics[width=0.7\linewidth,clip=true]{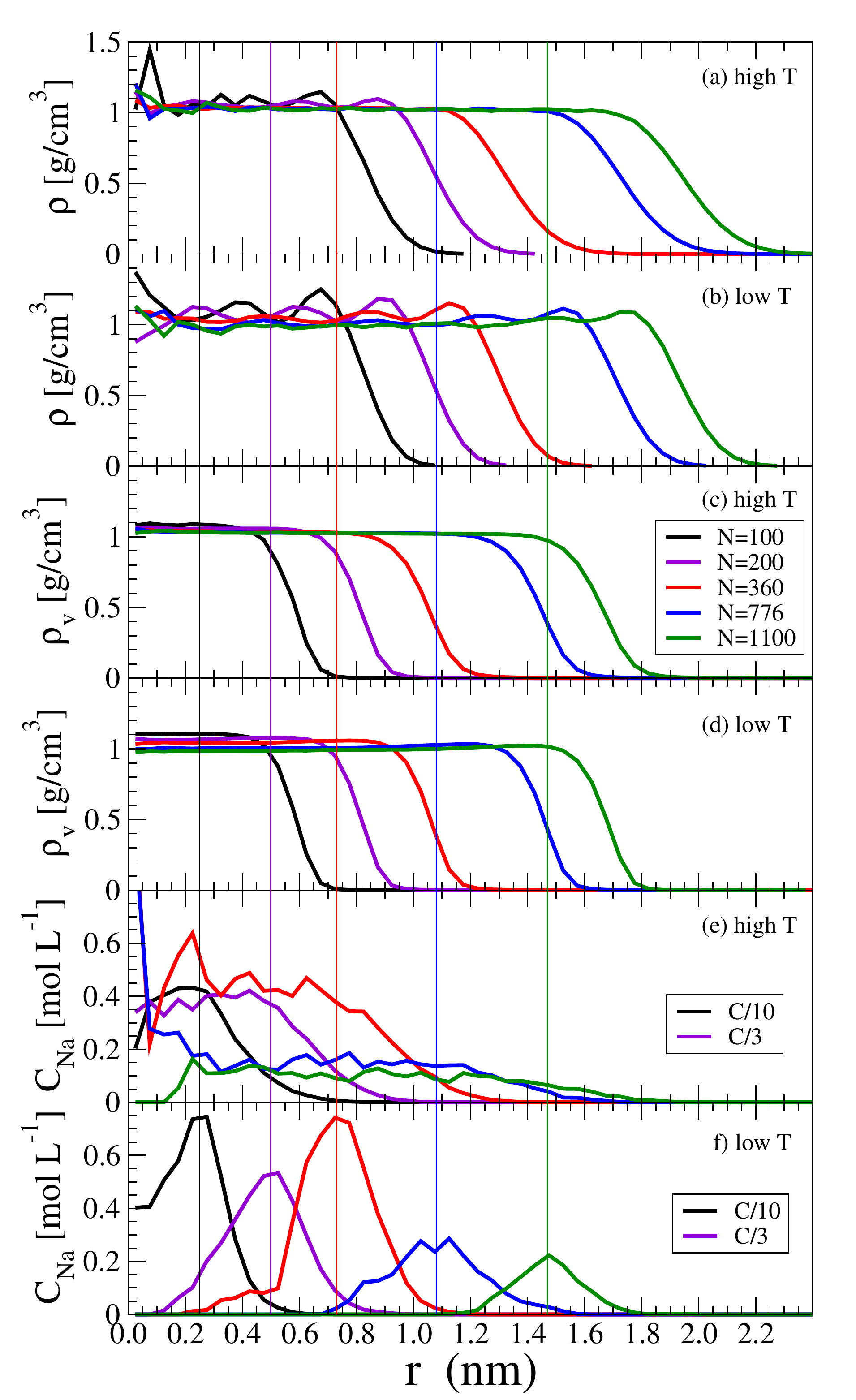}
    \caption{Pristine water nanodroplet structure and single \ce{Na} number density distributions for low temperature ($T=200$~K) and high temperatures ($T=260$~K for $N\le 200$ and $T=300$~K for $N\ge 360$).    
    Panels (a) and (b) show water density $\rho(r)$; panels (c) and (d) show water density based on molecular Voronoi volumes $\rho_V(r)$.  Panels (e) and (f) show \ce{Na+} concentration $C_{\rm Na}(r)$, which for $N=100$ and $N=200$ have been reduced by a factor of 10 and 3, respectively.
    }
    \label{fig:singleNa}
\end{figure}

Figure~\ref{fig:singleNa} shows $\rho(r)$ and $\rho_V(r)$ for pure water nanodroplets of all sizes studied at 
high (300~K for $N \ge 200$, 260~K for $N \le 200$) and low (200~K) temperature ($T$), 
with data taken from Ref.~\cite{maleknc2018}.  At high $T$, shown in Fig.~\ref{fig:singleNa}~(a), 
$\rho(r)$ is that of a typical liquid droplet, characterized by a flat (slowly decreasing) 
curve in the interior that decays sigmoidally to (near) zero over approximately an intermolecular 
distance at the liquid-vapor interface.  The exceptions are the curves for $N\le200$, that at $260$~K show 
some ordering or layering particularly near the surface. At low $T$, shown in Fig.~\ref{fig:singleNa}~(b), 
there are significant undulations in the density profiles for all nanodroplet sizes, and, 
as seen particularly well for the larger
nanodroplets, an increase in density as $r$ increases towards the surface.  

The undulations in $\rho(r)$ make it difficult to characterize how 
the local density changes with $r$, and for this reason we plot $\rho_V(r)$.  
Fig.~\ref{fig:singleNa}~(c) shows $\rho_V(r)$ monotonically decreasing (or flat) with $r$ for 
all nanodroplet sizes at high $T$.  $\rho_V(r)$ is significantly smoother than $\rho(r)$, as 
it does not depend on the number density itself, but rather on the Voronoi volume surrounding 
each water molecule.  An important feature of $\rho_V(r)$ is that it begins to decay to zero at 
approximately 0.3~nm, or an intermolecular distance, before $\rho(r)$; Voronoi volumes are very large, 
and Voronoi-based density very low, for molecules on the surface.  
Molecules falling within the range where $\rho_V(r)$ is high (near bulk values) are completely 
surrounded by other water molecules and are not on the surface.  
Surface molecules can be identified as those located where $\rho_V(r)$ is small, 
and molecules in the subsurface as those located an intermolecular distance below the point 
at which $\rho_V(r)$ has decayed to near zero.

Fig.~\ref{fig:singleNa}~(d) shows $\rho_V(r)$ for nanodroplets at 
low $T$. For $N\ge 200$, there is a significant increase in density in the subsurface layer.  
The density may well be higher for surface molecules, but $\rho_V(r)$ can not characterize this.  
This increased density at low $T$ appears to be a hallmark of cold water 
nanodroplets, and has not been reported for simple liquids to our knowledge.  
It is this heterogeneous environment in pure water that lends an interesting 
backdrop for studying ion distributions at low $T$.

 In Fig.~\ref{fig:singleNa}~(e), we plot $C_{\rm Na}(r)$ at high $T$ for systems 
 composed of a single \ce{Na+} ion within a nanodroplet.  
 Since the ion density is quite high for small nanodroplets, we divide $C_{\rm Na}(r)$ by 10
 and 3 for  $N=100$ and 200, respectively.  In all cases, the $C_{\rm Na}(r)$ is approximately
 constant in the interior of the droplet, and begins to decay within the subsurface,
 and decays to zero significantly before $\rho(r)$ does.  

Fig.~\ref{fig:singleNa}~(f) shows a dramatic difference in $C_{\rm Na}(r)$ at low $T$.  
Rather than being centered at $r=0$, the peak of $C_{\rm Na}(r)$ is located within 
0.1~nm of the peak in $\rho_V(r)$  (for $N\ge200$).  Thus, we see that in a nanodroplet 
with a heterogeneous radial density, as determined by $\rho_V(r)$, 
the single Na$^+$ ion tends to reside in the highest density environment. 

For $N=100$ at low $T$, $\rho_V(r)$ is approximately constant for $r<0.3$~nm, and then 
decreases with increasing $r$.   While a constant $C_V(r)$ for $r<0.3$~nm suggests 
that $C_{\rm Na}(r)$ should be uniform in this interior region, we see that $C_{\rm Na}(r)$ is in 
fact peaked just below 0.3~nm.  We do see, however, that the peak in 
$C_{\rm Na}(r)$ coincides with a local minimum in $\rho(r)$, suggesting that layering 
may play a significant role in determining where the Na ion resides in such small nanodroplets.


\begin{figure}
    \centering
    \includegraphics[width=0.7\linewidth,clip=true]{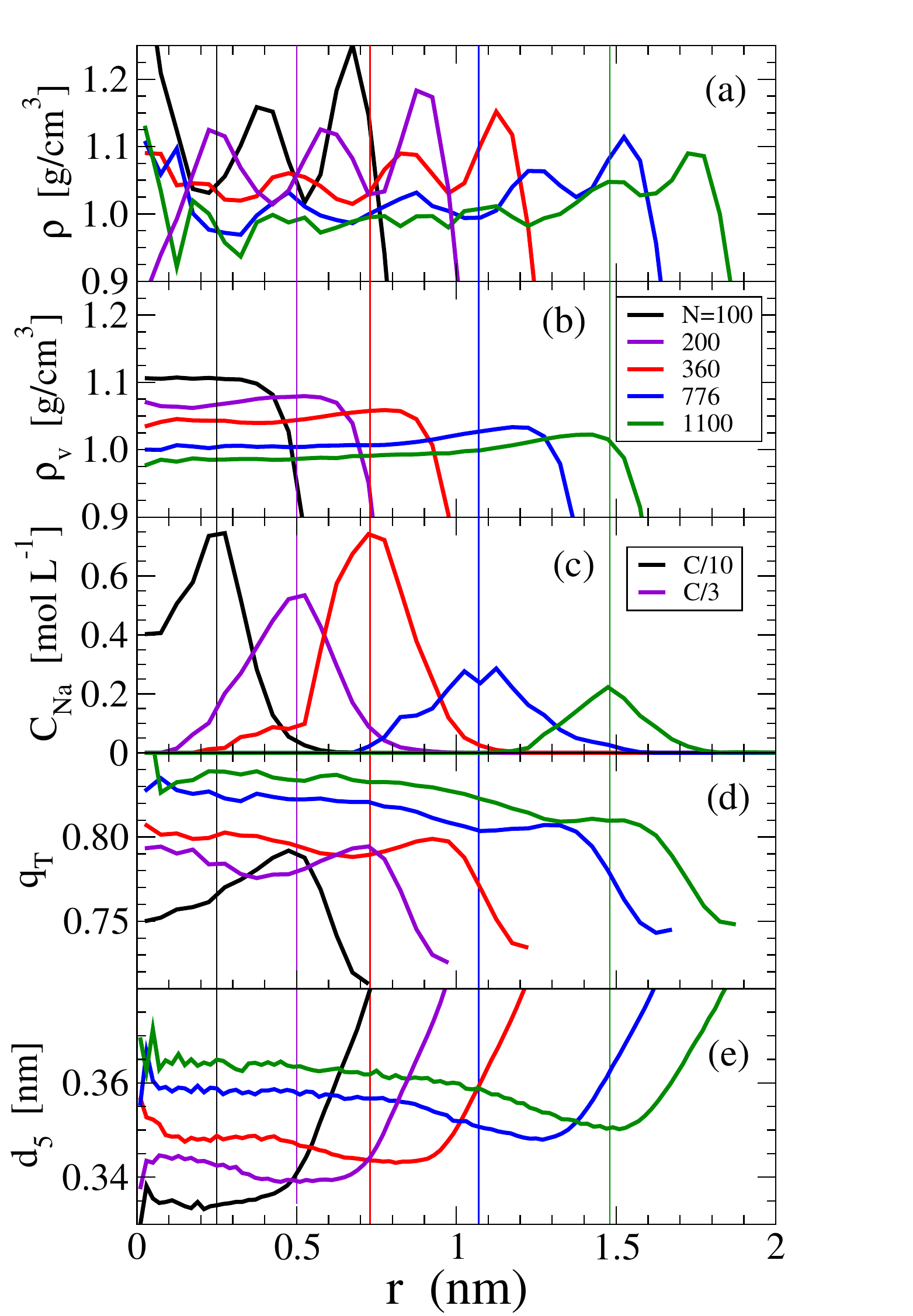}
    \caption{Single Na$^+$ concentration and measures of structure for 
      low temperature ($T=200$~K) for  nanodroplet sizes $N=100$ (black), 200 (violet), 360 (red), 776 (blue) and 1100 (green).  Shown are (a) $\rho$, (b)  $\rho_V$, (c) 
      C$_{\rm Na}$, (d) $q_T$ and (e) $d_5$, as functions or $r$. $\rho$, $\rho_V$, $q_T$ and $d_5$ are for pure water.  }  
    \label{fig:singleNaStruc}
\end{figure}

In Fig.~\ref{fig:singleNaStruc}, we plot for $200$~K $\rho(r)$, $C_{\rm Na}(r)$ for a single Na ion, $q_T(r)$ and $d_5(r)$, confirming that  for  $N=360$, 776, and 1100, the ion resides in a subsurface that is relatively disordered compared to the tetrahedral core. Data for $q_T(r)$ and $d_5(r)$ are taken from Ref.~\cite{malek2019}. $C_{\rm Na}(r)$ decays rapidly for increasing $r$ upon approaching the surface layer (where $d_5(r)$ rapidly increases) and for decreasing $r$ upon entering the region where $q_T(r)$ is high.  The exception is the $N=100$ nanodroplet, which does not have a tetrahedral core.  At this size, however, layering propagating from the surface extends to the droplet interior, and it is at a minimum in $\rho(r)$ that we find the peak in $C_{\rm Na}(r)$.  For the larger droplets too, it appears that the ion prefers to be in a trough, except for $N=1100$, where layering is relatively weak.

\clearpage

\section{S4. Convergence of the ion location and comparison of the radial probability density of various ions in a  776-\ce{H2O}-molecule droplet}

\begin{figure}
    \centering
    \includegraphics[width=\columnwidth]{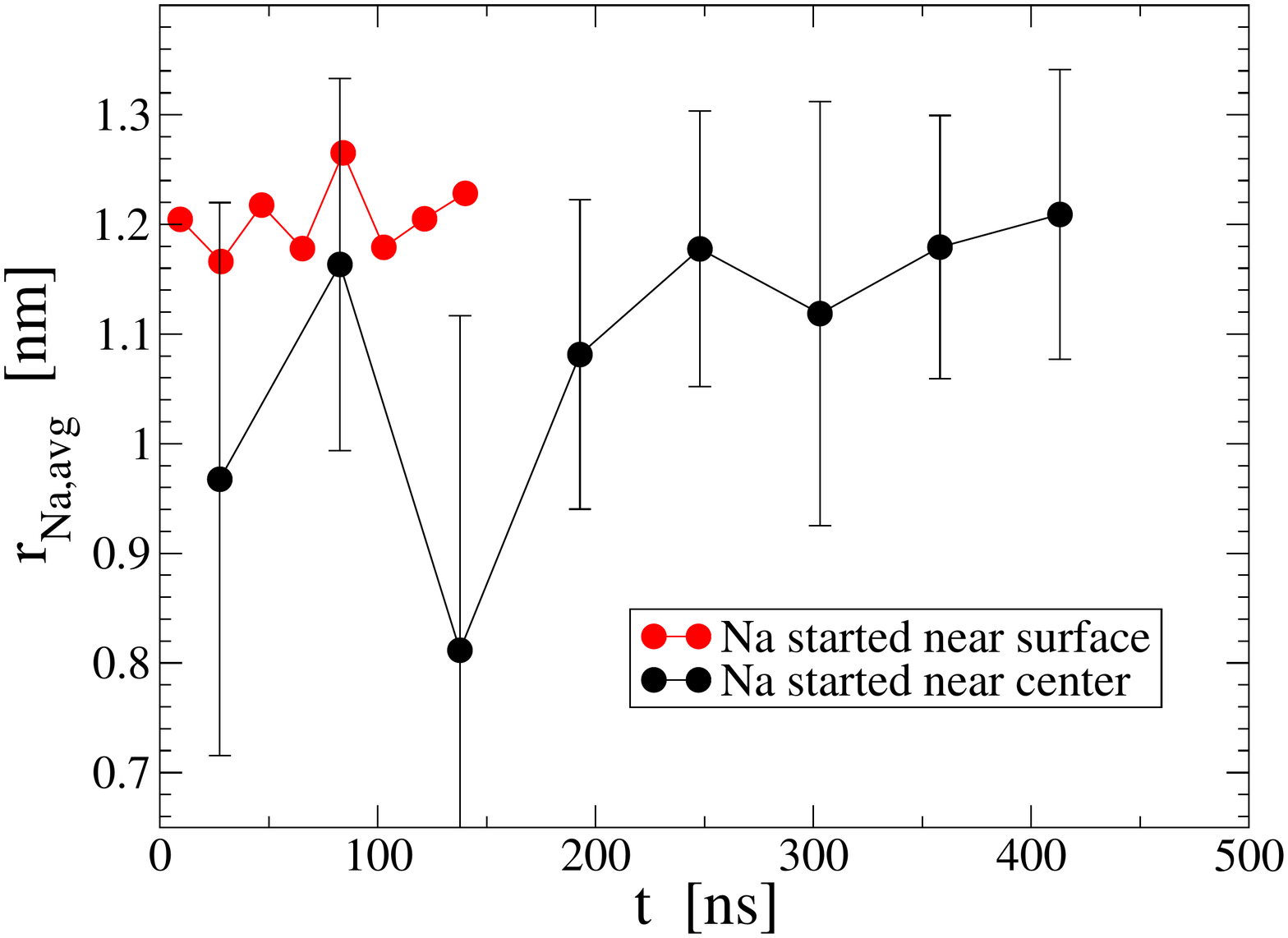}
    \caption{Average radial position of ion as a function of time in a
    system of a single \ce{Na+} ion and $N=776$ at $T=200$~K. Each time series is divided into eight equal blocks over which averages are reported.  Bars for the case where \ce{Na+} starts at the surface represent one standard deviation.  $R_B=0.73$~nm, $R_V=1.49$~nm and $R_e=1.77$~nm for this case.}
    \label{fig:NaN776converge}
\end{figure}

\newpage

\begin{figure}
\centering
\begin{minipage}{.43\linewidth}
  \includegraphics[width=\linewidth]{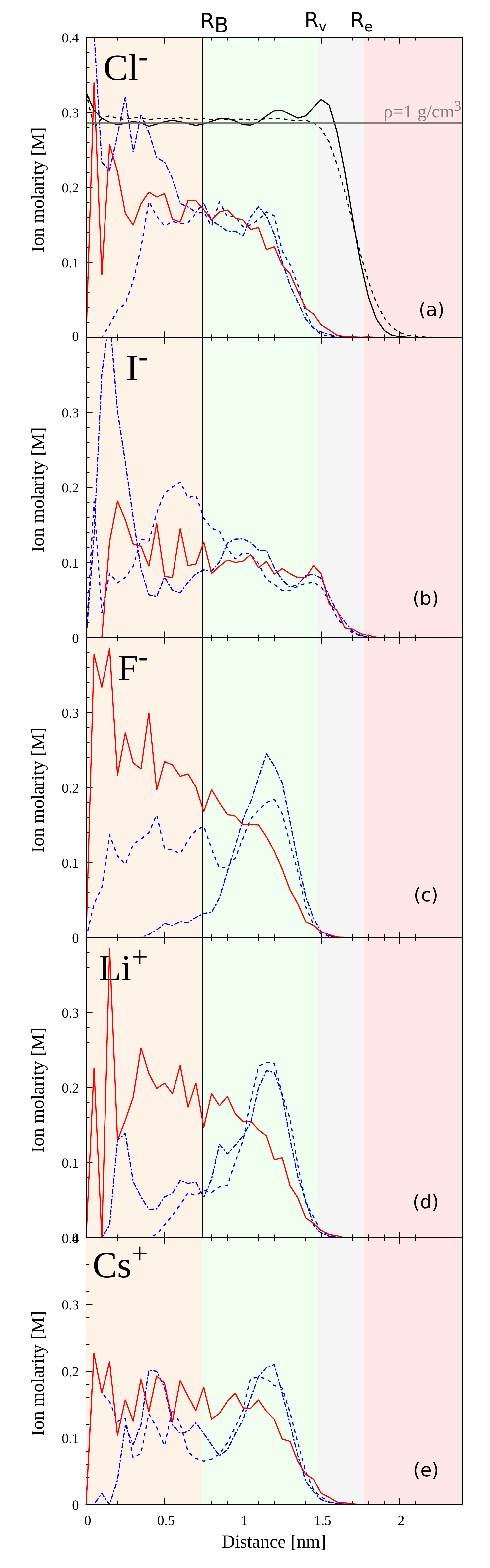}
  \captionof{figure}{Concentration profiles for various ions. Red line at 300~K, blue lines at 200~K.}
  \label{fig:comparison-distr}
\end{minipage}
\hspace{.05\linewidth}
\begin{minipage}{.43\linewidth}
  \includegraphics[width=\linewidth]{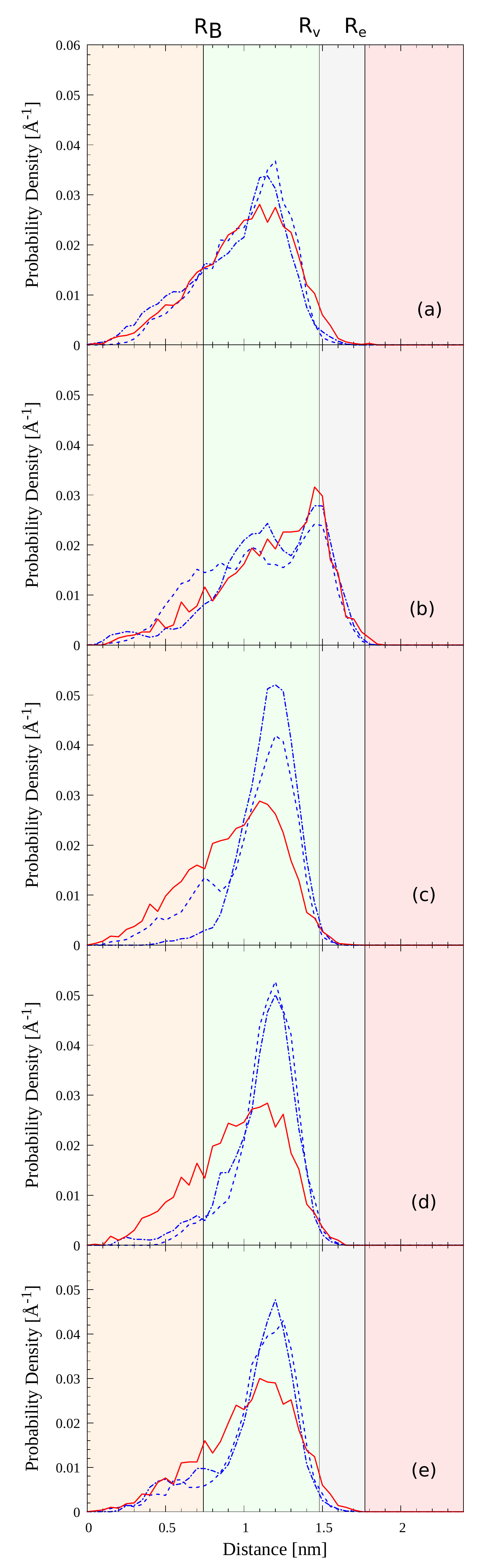}
  \captionof{figure}{Same as Fig.~\ref{fig:comparisons-raw} but with no volume normalization.}
  \label{fig:comparisons-raw}
\end{minipage}
\end{figure}

\newpage

\begin{figure}
    \centering
    \includegraphics[width=1.2\columnwidth]{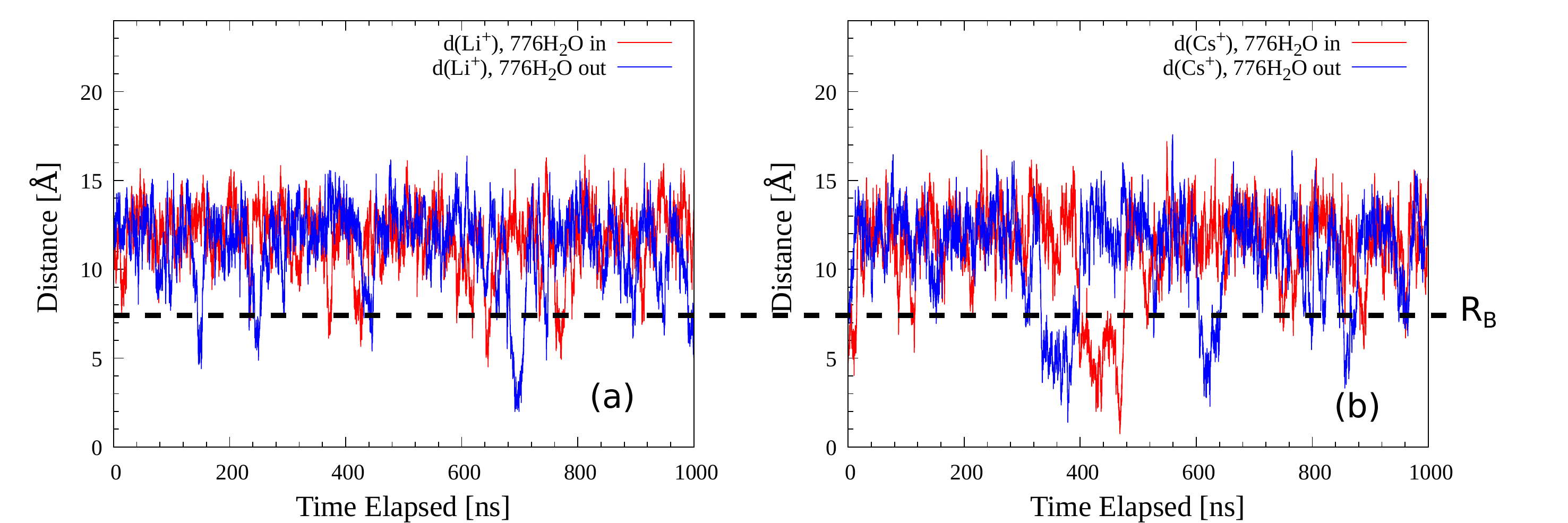}
    \caption{Time evolution of the ion, \ce{Cs+} and \ce{Li+}, distance from the droplet COM. The two lines (blue
  and red) correspond to the different MD trajectories starting from the droplet center and surface. The data
  presented here and analyzed in Figs.~\ref{fig:comparison-distr} and \ref{fig:comparisons-raw} are those following
200 ns of initial equilibration time.}
    \label{fig:time-Cs-Li}
\end{figure}

\begin{figure}
    \centering
    \includegraphics[width=1.2\columnwidth]{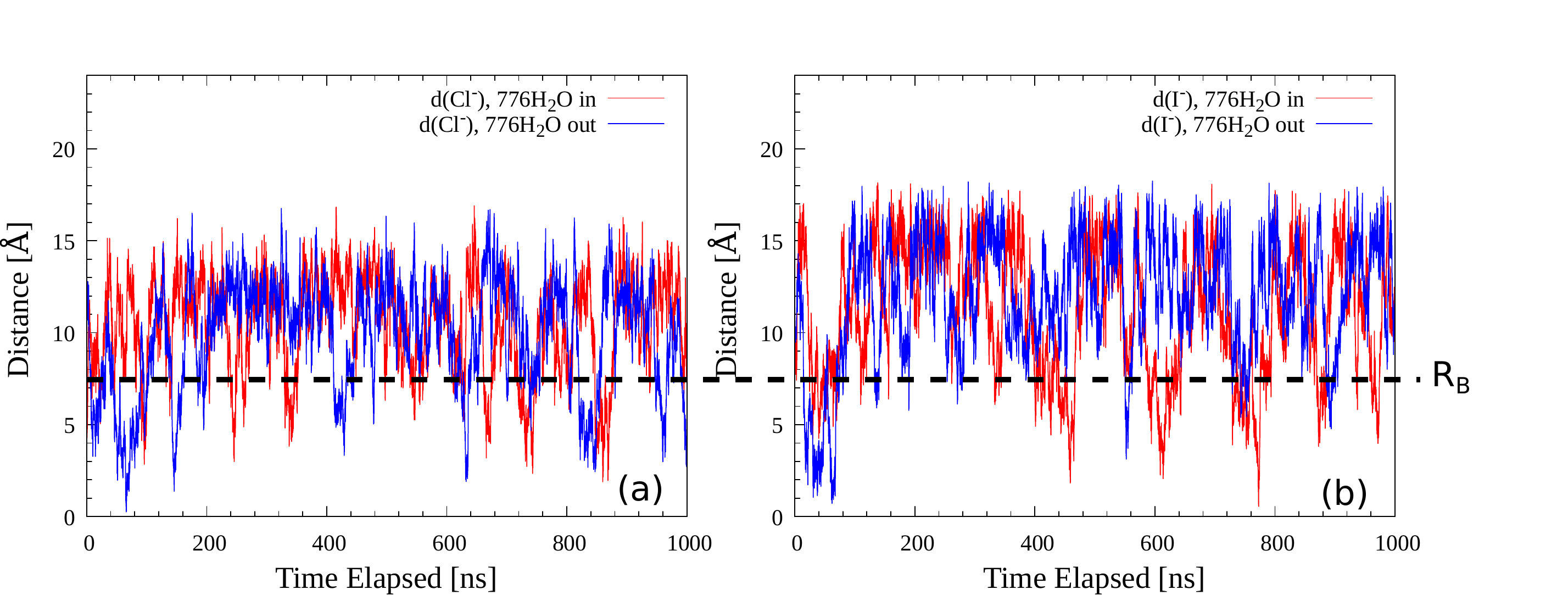}
    \caption{Same as Fig.~\ref{fig:time-Cs-Li} but for \ce{Cl-} and \ce{I-}.}
    \label{fig:time-Cl-I}
\end{figure}

\begin{figure}
    \centering
    \includegraphics[width=\columnwidth]{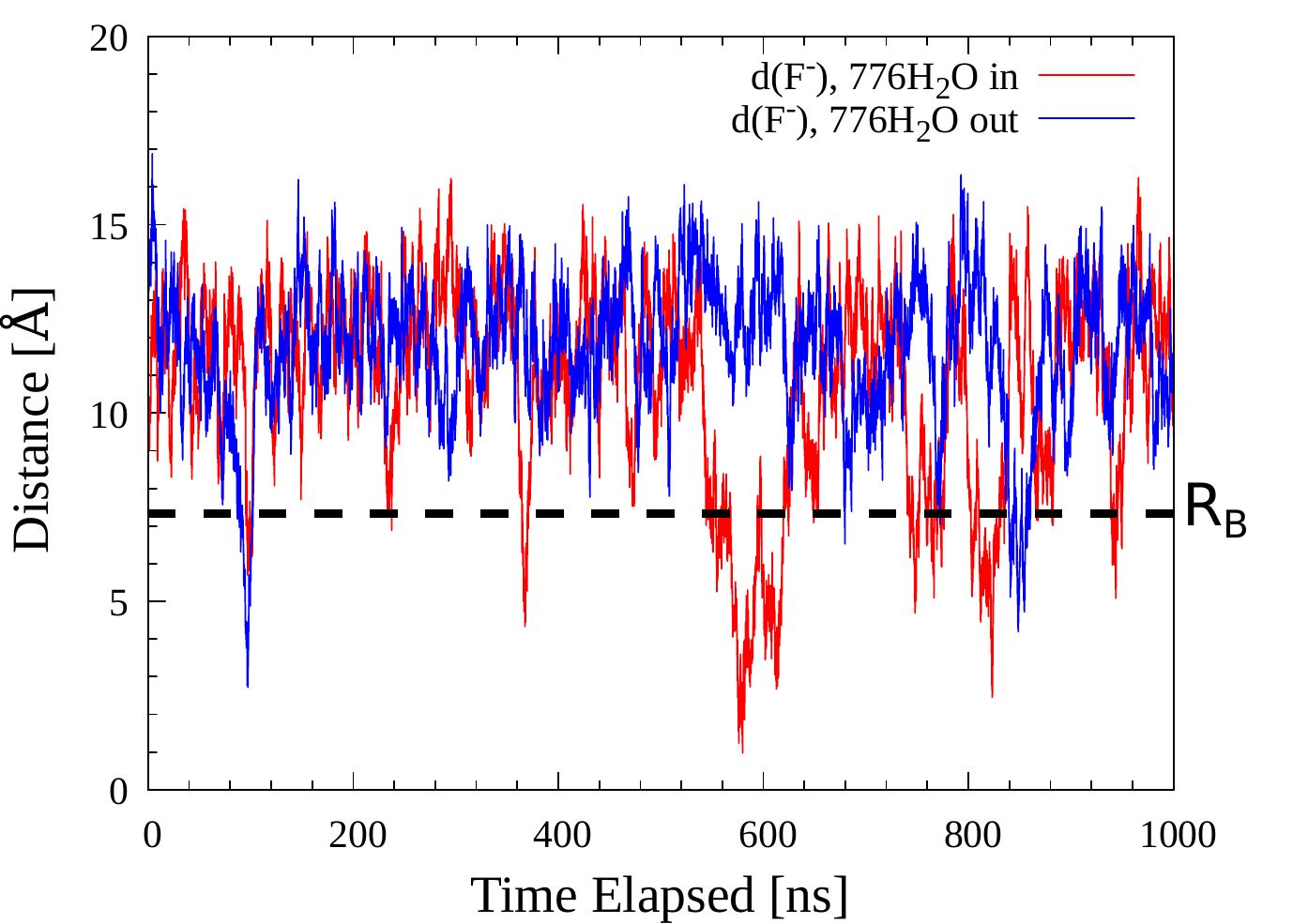}
    \caption{Same as Fig.~\ref{fig:time-Cs-Li} but for \ce{F-}.}
    \label{fig:time-F}
\end{figure}

\begin{figure}
    \centering
    \includegraphics[width=\columnwidth]{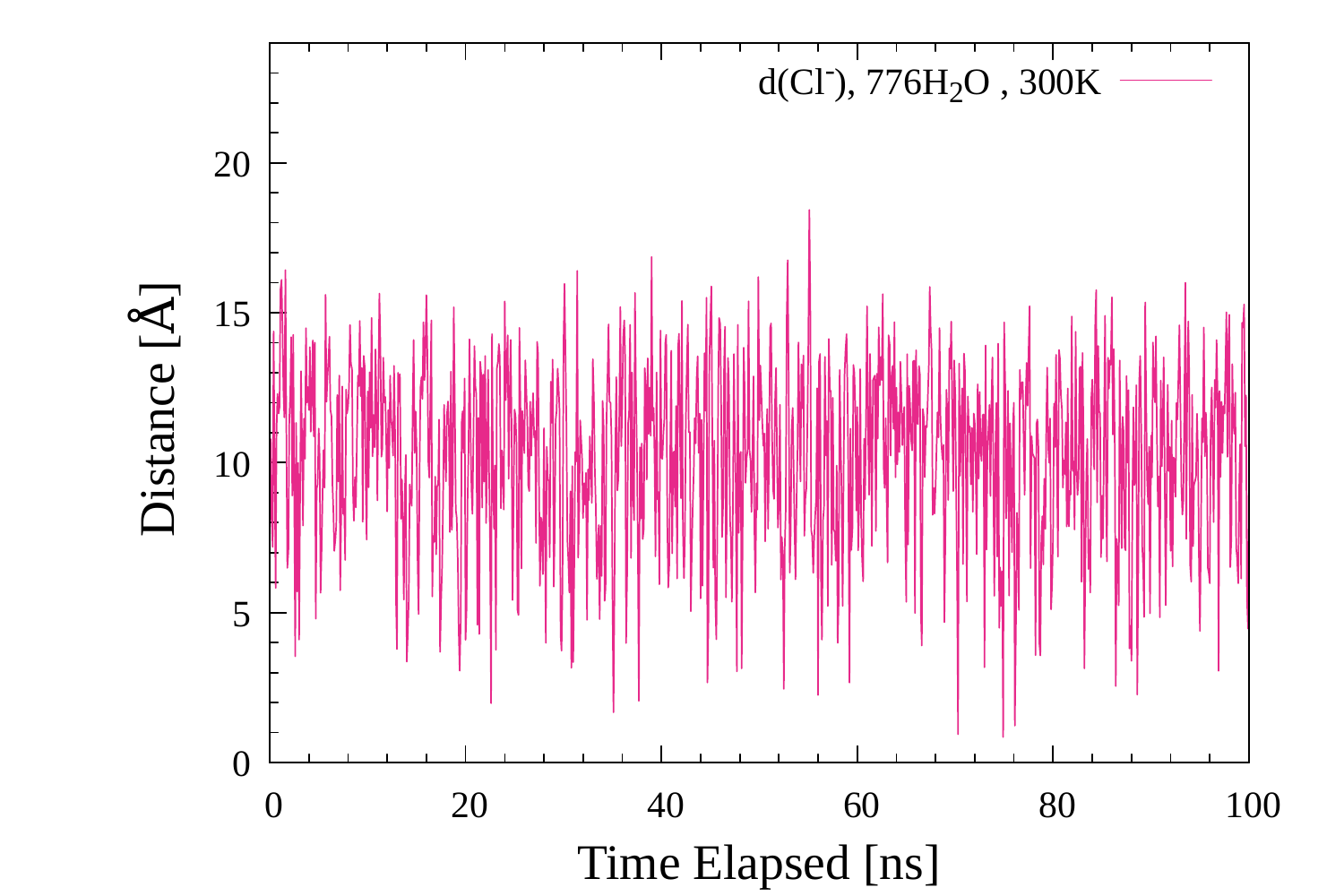}
    \caption{Same as Fig.~\ref{fig:time-Cs-Li} but for \ce{Cl-} at 300~K.}
    \label{fig:time-Cl-300K}
\end{figure}

\clearpage

\section{S5. \ce{Cl-} and \ce{F-} radial distribution functions in a 776-\ce{H2O}-molecule droplet}
\begin{figure}
\centering
\begin{subfigure}[htbp]{0.5\textwidth}
\centering
\includegraphics[width=\textwidth]{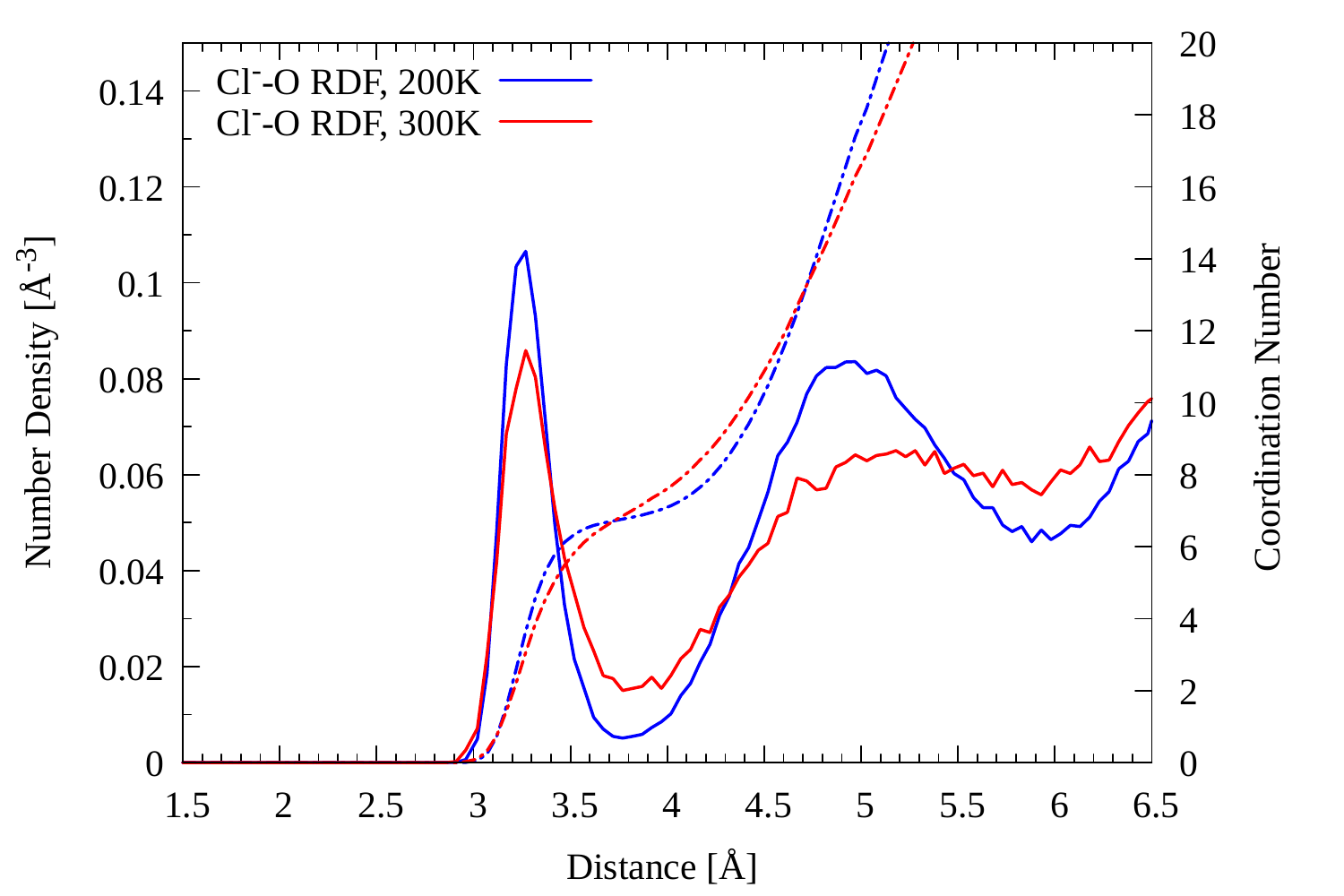}
\caption{}
\end{subfigure}
\begin{subfigure}[htbp]{0.5\textwidth}
\centering
\includegraphics[width=\textwidth]{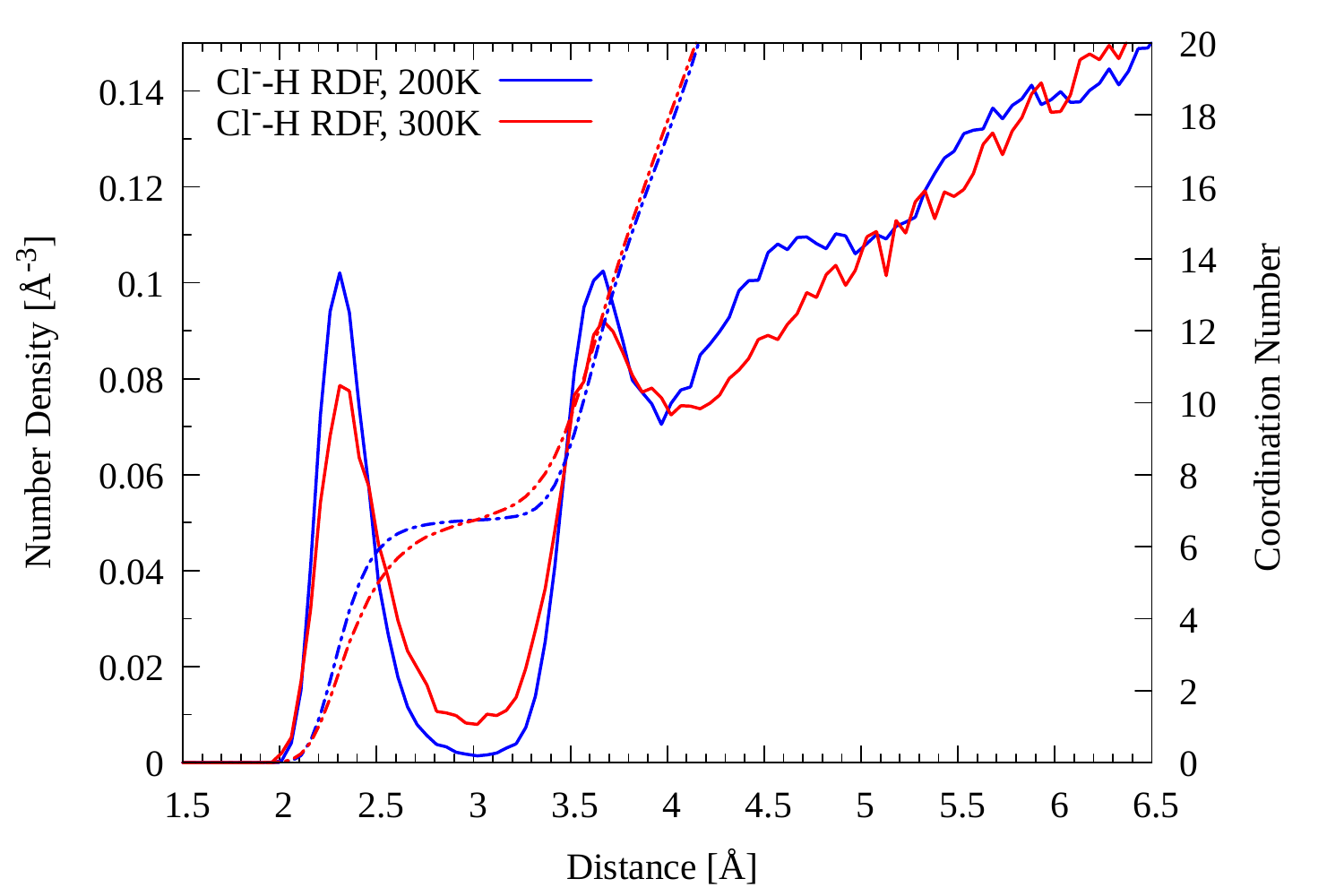}
\caption{}
\end{subfigure}

\caption{Radial distribution function ($g(r)$) at 200~K and 300~K for (a) O-Cl and (b) H-Cl.}
\label{fig:RDF-CL}
\end{figure}

\newpage

\begin{figure}
\centering
\begin{subfigure}[htbp]{0.5\textwidth}
\centering
\includegraphics[width=\textwidth]{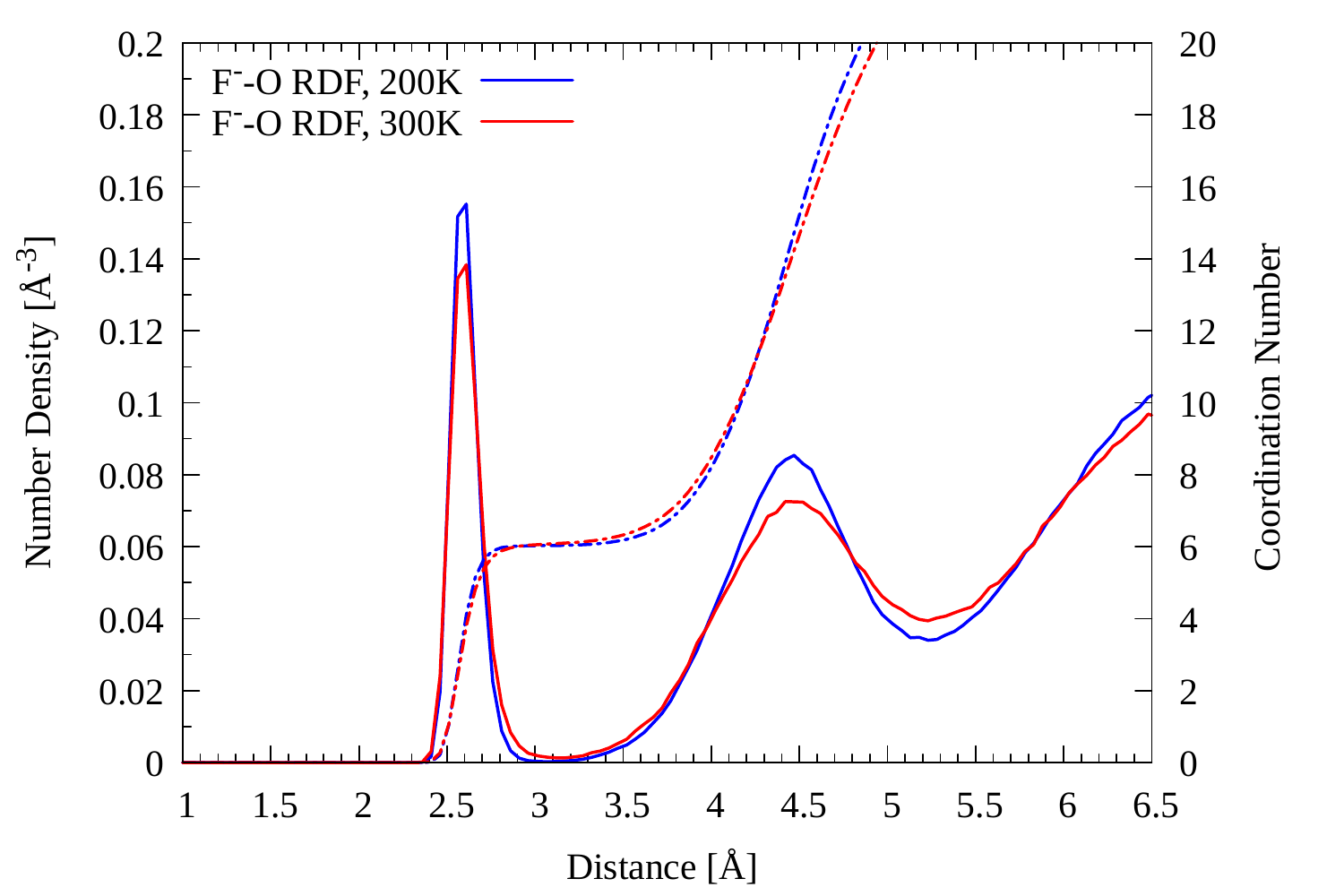}
\caption{}
\end{subfigure}
\begin{subfigure}[htbp]{0.5\textwidth}
\centering
\includegraphics[width=\textwidth]{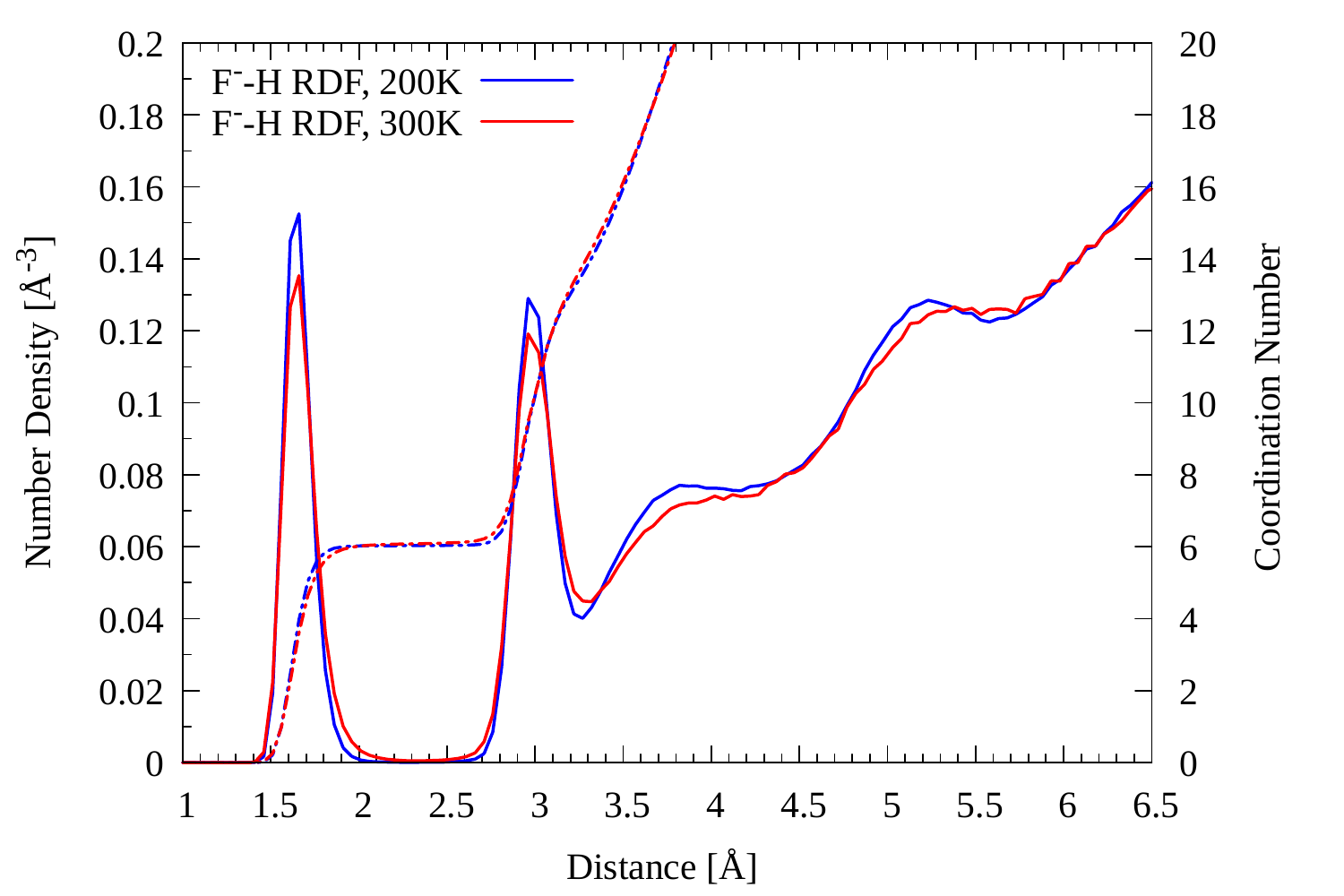}
\caption{}
\end{subfigure}

\caption{Radial distribution function ($g(r)$) at 200~K and 300~K for (a) O-F and (b) H-F.}
\label{fig:RDF-F}
\end{figure}

\newpage

\clearpage

\section{S6. Radial distributions of ions using polarizable force field} 
\begin{figure}
    \centering
    \includegraphics[width=0.7\linewidth,clip=true]{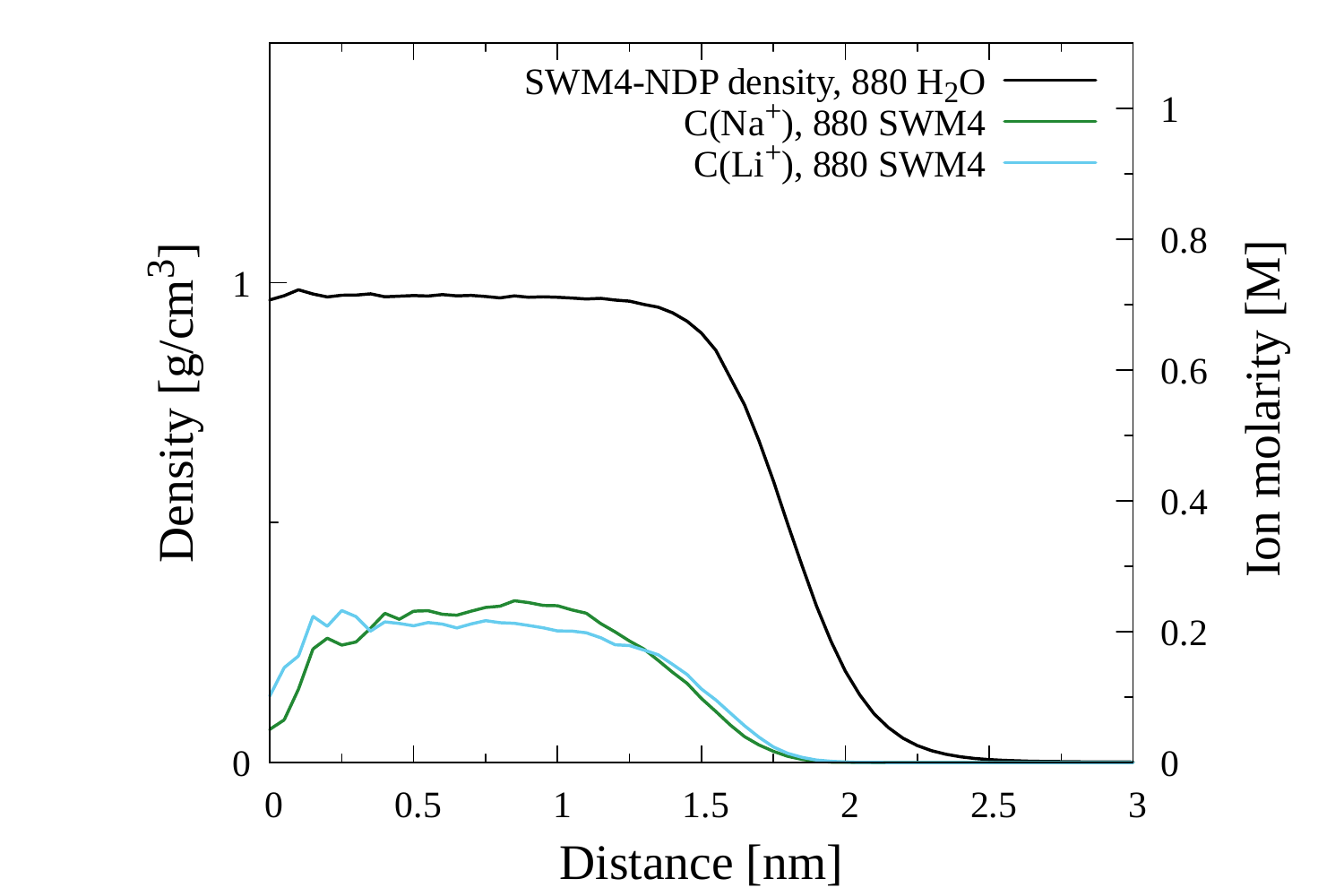}
    \caption{Concentration profiles of \ce{Li+} and  \ce{Na+} ions (measured in the right y-axis) and water density 
    (measured in the left y-axis) in a droplet comprising 880 \ce{H2O} molecules at $T=350$~K.} 
    \label{fig:polarizable-T350K}
\end{figure}

\begin{figure}
    \centering
    \includegraphics[width=0.7\linewidth,clip=true]{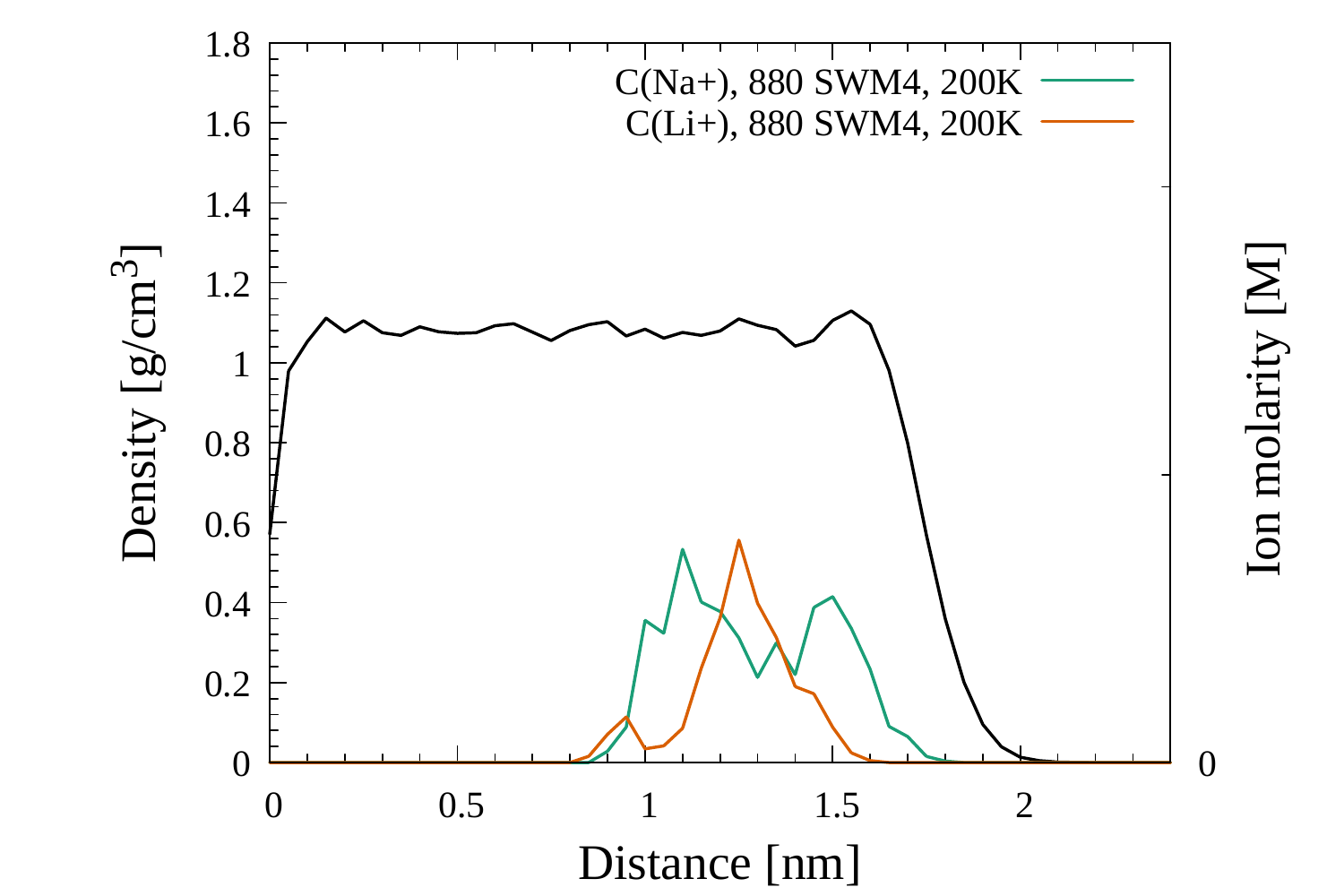}
    \caption{Same as Fig.~\ref{fig:polarizable-T350K} but at $T=200$~K.}
    \label{fig:polarizable-200K}
\end{figure}

\clearpage

\section{S7. Radial distribution of multiple \ce{Na+} ions in a droplet of 776 \ce{H2O} molecules}

\begin{figure}
    \centering
    \includegraphics[width=0.7\linewidth,clip=true]{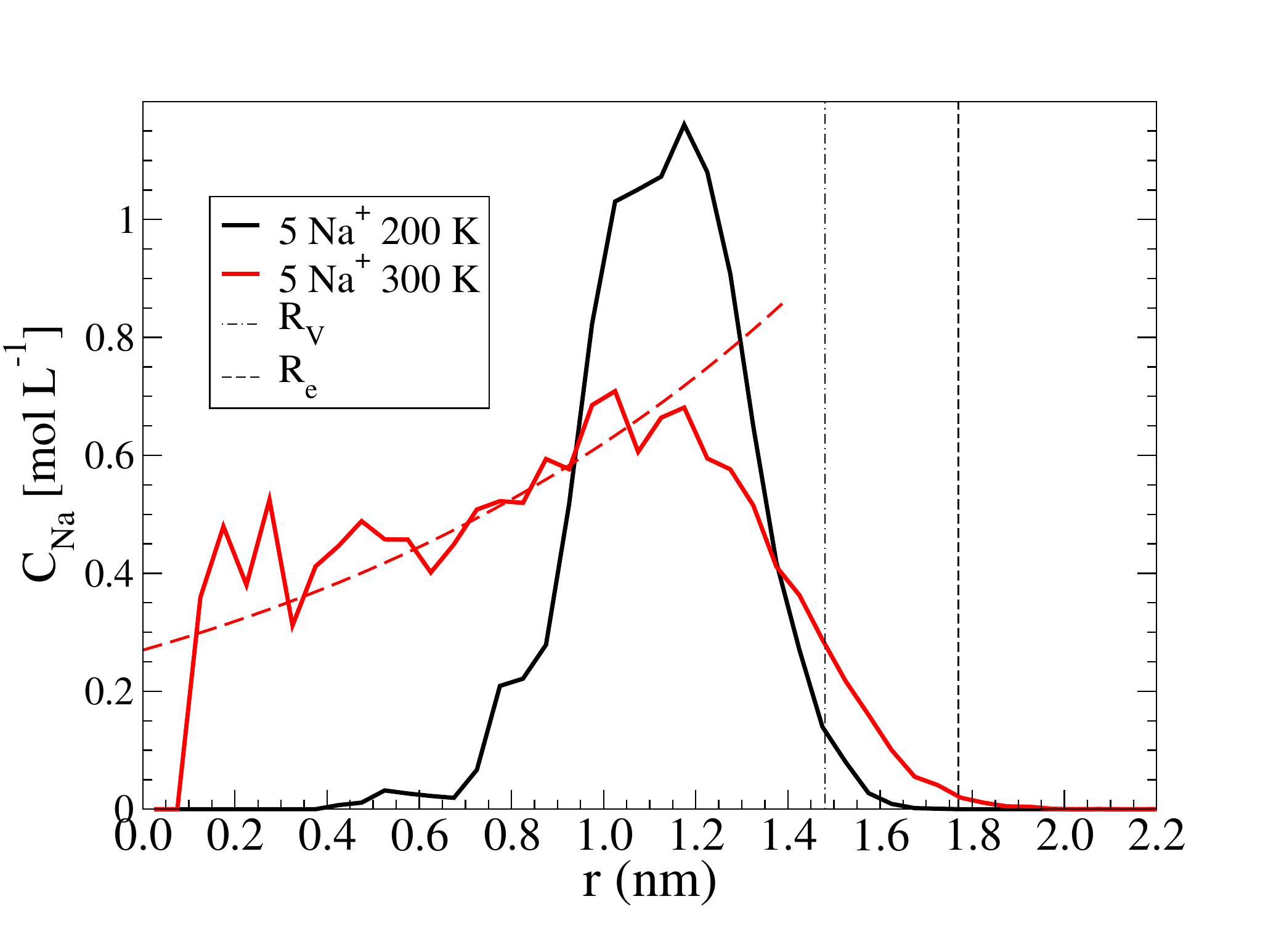}
    \caption{C$_{\rm Na}(r)$ for the 776-molecule droplet with 5 \ce{Na+} ions.
    Red dashed curve indicates a fit 
      to Eq.~\ref{eq:2} (main text). The distribution at 300~K
decays (toward the droplet's COM) as an exponential function fitted
by $0.42 \exp(-(1.14-r)/1.2)\times 10/6.022$, where $\lambda_{\mathrm{PB}} \approx 1.2$~nm. The vertical lines mark $R_V$ and $R_e$.}
    \label{fig:multiNaStruc}
\end{figure}

\newpage

\providecommand{\latin}[1]{#1}
\makeatletter
\providecommand{\doi}
  {\begingroup\let\do\@makeother\dospecials
  \catcode`\{=1 \catcode`\}=2 \doi@aux}
\providecommand{\doi@aux}[1]{\endgroup\texttt{#1}}
\makeatother
\providecommand*\mcitethebibliography{\thebibliography}
\csname @ifundefined\endcsname{endmcitethebibliography}
  {\let\endmcitethebibliography\endthebibliography}{}